\documentclass[useAMS,usenatbib,usegraphicx]{mn2e}
\title[Galactic resonance rings]
{Galactic resonance rings. Modelling of motions in the wide solar
neighborhood}
\author[A.M. Melnik ]{A.M.Melnik\thanks{E-mail: anna@sai.msu.ru}
\\ Sternberg Astronomical Institute, Lomonosov Moscow State
University, Universitetskii pr. 13, Moscow, 119991 Russia}

\begin{document}

\date{Accepted 2017 December 00. Received 2017 December 00; in original form 2017 December 00}


\maketitle

\label{firstpage}

\begin{abstract}
Models of the Galaxy with analytical Ferrers bars can  reproduce
the residual velocities of OB-associations in the Sagittarius,
Perseus and Local System stellar-gas complexes located within 3
kpc solar neighborhood.  Ferrers ellipsoids with density
distribution defined by power indices $n=1$ and 2 are considered.
The success in the reproduction of the velocity in the Local
System is due to the large velocity dispersion which weakens the
resonance effects by producing smaller systematic motions. Model
galaxies form nuclear,  inner and outer resonance rings $R_1$ and
$R_2$.  The outer rings  $R_2$ manage to catch twice more
particles  than  the rings $R_1$. The outer Lindblad resonance of
the bar (OLR) is located  0.4 kpc beyond the solar circle,
$R_{OLR}=R_0+0.4$ kpc, corresponding to  the bar angular velocity
of $\Omega_b=50$ km s$^{-1}$ kpc$^{-1}$. The solar position angle
with respect to the bar, $\theta_b$, providing the agreement
between model and observed velocities is 40--52$^\circ$.
Unfortunately, models considered cannot reproduce the residual
velocities in the Carina and Cygnus stellar-gas complexes. The
redistribution of the specific angular momentum, $L$, is found
near the Lindblad resonances of the bar (ILR and OLR): the average
value of $L$ increases (decreases)  at the radii a bit smaller
(larger) than those of the resonances that can be connected with
the existence of two types of periodic orbits elongated
perpendicular to each other there.

\end{abstract}


\section{Introduction}

The presence of a bar in the Galaxy is the signpost of secular
evolution of galaxy structure \citep{kormendy2004}. After the end
of the epoch of violent galaxy-galaxy interactions ($\sim 7$ Gyr
ago) secular processes causes the bar formation in disc galaxies.
Observations suggest that the fraction of disc galaxies containing
a bar decreases towards higher redshifts and that most massive
galaxies form bars much earlier  than lower mass ones
\citep{sheth2008,melvin2013}. Modelling shows that time-scale over
which a bar forms increases strongly with decreasing disc-to-total
mass fraction \citep[e.g.][]{athanassoula1986,fujii2018}. Though
bars can  spontaneously form  in dynamically cold discs
\citep{ostriker1973}, the bar fraction depends on the environment:
in disk dominated galaxies  tidal interactions can trigger bar
formation \citep[e.g.][]{elmegreen1990, mendez2012, martinez2017}.
The rigid rotation of the bar in the differentially rotating disc
causes the appearance of the resonances and the formation of
resonance rings \citep{buta2017}.

There is a lot of  evidence that our Galaxy includes a bar.
Infrared observations of the inner Galactic plane
\citep{dwek1995,benjamin2005,cabrera-lavers2007, churchwell2009,
gonzalez2012}, gas kinematics in the inner Galaxy
\citep{pohl2008,gerhard2011},  X-shaped distribution of red giants
in the central region derived from  BRAVA, WISE and VVV data
\citep{li2012,ness2016,simion2017} confirm the presence of the bar
in the Galaxy. The estimates of the length of the bar major
semi-axis lie in the range 3--5 kpc which corresponds to the bar
angular velocity 40--70 km s$^{-1}$.

The resonance between the frequency of  orbital rotation  with
respect to the bar  and the frequency of epicyclic motions causes
the formation of the elliptical resonance rings \citep{buta1995,
buta1996}. The condition of the resonance is following:

\begin{equation}
\frac{n}{m}= \frac{\kappa}{\Omega-\Omega_b}, \label{resonance}
\end{equation}

\noindent where  $n$ shows the number of full epicyclic
revolutions which are made by a star  rotating on a circular orbit
around the galactic center during $m$ orbital revolution with
respect to the bar. Usually the case of $m=1$ is considered. The
fraction $n/m=\pm2/1$ corresponds to the Inner Lindblad Resonance
(ILR, $+2/1$) and the  Outer Lindblad Resonance (OLR, $-2/1$)
besides the high order resonances $\pm$4/1 are also important
\citep{athanassoula1992,contopoulos1989}.

Modelling of the resonance rings shows  that the  outer rings are
forming  near the OLR of the bar while inner and nuclear rings are
emerging near the  inner 4/1 resonance and the ILR, respectively
\citep{schwarz1981, byrd1994,rautiainen1999,
rautiainen2000,rodriguez2008,pettitt2014,li2015,sormani2018}.

The outer rings have two preferable orientations with respect to
the bar: the rings $R_1$ are elongated perpendicular to the bar
while the rings $R_2$ are stretched along the bar. Of two outer
rings, $R_1$ lies a bit closer to the galactic center than $R_2$.
Some of them don't have pure elliptical shape but include a break
so that they rather resemble two tightly wound spiral arms. Broken
rings are named pseudorings and are marked with apostrophe, for
example, $R_1'$ and $R_2'$ \citep{buta1995,buta1996,buta1991}.

All resonance rings  are supported by  main periodic orbits.   The
main periodic orbits are stable orbits close to the circular ones
in the unperturbed case.  Such orbits  are followed by a large set
of quasi-periodic orbits. There are two basic families of stable
direct periodic orbits, $x_1$ and $x_2$. The family $x_2$ of
stable periodic orbits exists only between two ILRs. There is also
a third   family of periodic orbits, $x_3$, which consists of
unstable orbits. The main periodic orbits $x_1$ inside the
corotation radius (CR) are elongated along the bar and form the
backbone of the bar. The $x_2$ orbits are  elongated perpendicular
to the bar and support the nuclear rings.  Near the OLR of the bar
the main family of periodic orbits $x_1$ is splitting into two
families: $x_1(1)$ and $x_1(2)$. The main stable periodic orbits
$x_1(2)$ lying between the  $-4/1$ and $-2/1$ (OLR) resonances are
elongated perpendicular to the bar while  orbits $x_1(1)$ located
outside the OLR are stretched along the bar. Periodic orbits
$x_1(2)$ support outer rings $R_1$ while orbits $x_1(1)$ support
outer rings $R_2$. \citep{contopoulos1980, contopoulos1989,
schwarz1981, buta1996}.

Study of invariant manifolds associated with unstable periodic
orbits around  Lagrangian equilibrium points $L_1$ and $L_2$ shows
that they  can also give rise to   spiral-like and ring-like
structures in barred galaxies \citep{romero-gomez2007,
athanassoula2009,jung2016}.

Analysis of the mid-infrared images of galaxies detected by the
Spitzer Space Telescope \citep{sheth2010} reveals that the
fraction of galaxies hosting outer rings or preudorings increases
with increasing bar strength from 15 (SA) to  32 per cent
(SA\underline{B}) and then drops to 20 per cent for stronger bars
(SB) \citep{comeron2014}.

\citet{elmegreen1985}  discover two types of bars: large bars with
nearly a constant surface brightness mostly found in early type
galaxies and smaller bars with nearly exponential profiles mainly
observed in late-type galaxies. \citet{laurikainen2005} show that
"flat" bars in early-type galaxies can be described well either by
a S\'ersic or a Ferrers function.

The presence of the outer rings in the Galaxy was first suggested
by \citet{kalnajs1991}. The main advantage of models with outer
rings is that they don't need spiral-like potential perturbation
to create long-lived elliptical structures at the galactic
periphery. Outer rings are forming in 200-500 Myr after the bar
formation and can exist for several Gyrs \citep{rautiainen2000,
rautiainen2010}.

The angle between the major axis of the bar and the Sun--Galactic
center line or the so-called solar position angle with respect to
the bar, $\theta_b$,  derived from   data of the infrared surveys
(GLIMPSE, 2MASS, and VVV) has the value  of 40--45$^\circ$ so that
the end of the bar closest to the Sun is located in the first
quadrant \citep{benjamin2005, cabrera-lavers2007, gonzalez2012}.
Besides, a reconstruction of the Galactic CO maps with smoothed
particle hydrodynamics gives the best results for the solar
position angle of $\theta_b\approx 45^\circ$ \citep{pettitt2014}.

\citet{melnikrautiainen2009} using  models with analytical Ferrers
bars study the Galactic kinematics in the 3 kpc solar
neighborhood. Their models form the two-component outer rings
$R_1R_2'$ after  $\sim800$ Myr from the start of simulation. The
gas subsystem includes 5 10$^4$ massless gas particles which can
collide with each other inelastically.  The best agreement between
model and observed velocities corresponds to the solar position
angle  $\theta_b=45\pm5^\circ$. These models can reproduce the
average velocities of OB-associations in the Perseus and
Sagittarius stellar-gas complexes but failed in the Local System,
Cygnus and Carina stellar-gas complexes.

The fact that the position angle of the Sun with respect to the
bar, $\theta_b$, is close to 45$^\circ$ means that the 3-kpc solar
neighbor can harbor both a segment of the outer ring $R_1$ and a
segment of the ring $R_2$. The study of the distribution of
classical Cepheids and young open clusters reveals the existence
of "the tuning-fork-like" structure which can be interpreted as
two segments of the outer rings fusing together near the Carina
stellar-gas complex \citep{melnik2015, melnik2016}. Note also that
models with the two-component outer ring $R_1R_2'$ can explain the
position of the Sagittarius-Carina arm in the Galactic disc: a
segment of the ring $R_1$ outlines the Sagittarius arm while an
arch of the outer ring $R_2$ lies in the vicinity of the Carina
arm \citep{melnik2011}.

\citet{rautiainen2010} build  N-body models of the Galaxy which
demonstrate the development of a bar and the formation of the
outer rings which  being formed  persist till the end of
simulation (6 Gyr). The special feature of N-body models is fast
changes of velocities of model particles which can be separated
into quick stochastic changes due to irregular forces and
quasi-periodic slow oscillations due to slow modes (patterns
rotating slower than the bar). Thus, the averaging of model
velocities over a large time interval is required for a comparison
with  observed velocities. In  N-body models by
\citet{rautiainen2010}, the velocities of model particles are
averaged over the time interval of 1 Gyr in the reference system
corotating with the bar. The averaged model velocities appear to
be able to reproduce the observed velocities in the Sagittarius,
Perseus and Local System stellar-gas complexes. The advantage of
models without self-gravity in kinematical studies is that they
enable us to compare observe and model velocities directly without
averaging.

Models presented here don't include spiral arms because both
resonance rings and spiral arms are invoking to explain the same
things: the systematic velocity deviations from the rotation curve
and the increased density of young objects in some regions. Any
travelling spiral density wave \citep{lin1964} winds up around the
Lindblad resonances after a few time revolution periods
\citep{toomre1969}. The mechanism (WASER) including the reflection
of the travelling wave in the central region can support a steady
spiral pattern \citep{mark1976,bertin1996} but it gives small
amplification to support the shock fronts as well
\citep{athanassoula1984,binney1987}. The shock fronts forming in
spiral arms due to collisions in gas subsystem act in the same
direction causing the drift of gas from the CR  towards the
Lindblad resonances \citep{toomre1977}. Another conception of the
galactic spiral structure suggests short transient  spiral arms
forming in self-gravitating galactic discs due to swing
amplification mechanism  \citep{julian1966, toomre1981}. This
mechanism can be very powerful and transient ragged spiral arms
often appear in simulations with live discs \citep[][and other
papers]{pettitt2015, baba2009,baba2013,grand2012,d_onghia2013}.
However, the pitch angle of these  short-lived arms is quite
large, $i=20\textrm{--}30^\circ$. The theoretical prediction of
its value is  $i=24^\circ$ \citep{melnik2013,michikoshi2014}. But
the Galactic global spiral arms seem to have a considerably
smaller pitch angle, $10\textrm{--}15^\circ$ \citep[for
example,][and references therein]{georgelin1976,
russeil2003,vallee2015}. Generally, the conception of the Galactic
spiral arms has a lot of difficulties. Nevertheless, many
kinematical and morphological features of the Galaxy  can be
explained in terms of spiral arms \citep[][and other
papers]{rastorguev2017,
bobylev2018,xu2018,grosbol2018,antoja2018,kawata2018,ramos2018}.

In this paper I  present  several models with analytical Ferrers
bars which can reproduce  the observed velocities in the
Sagittarius, Perseus and Local System stellar-gas complexes. The
crucial factor which determines the success in the Local System
appears to be a  large  velocity dispersion which weakens the
resonance effects. Section 2 considers the distribution of
observed velocities of young stars in the Galactic disc; Section 3
describes  the models; Section 4  compares  model and observed
velocities, studies the  distribution of the surface density,
velocity dispersions and the angular momentum along the radius;
Section 5 infers the main conclusions.

\section{The observed velocities of OB-associations in the 3-kpc solar neighborhood}

The  velocities of OB-associations give the most reliable
information about the distribution of velocities of young objects
in a wide solar neighborhood. The catalog by
\citet{blahahumphreys1989} includes  91 OB-associations, $\sim85$
per cent of which include at least one star of spectral types
earlier than B0, whose age is supposed to be less than 10 Myr
\citep{bressan2012}, so the average velocities of OB-associations
must be very close to the velocities of their parent giant
molecular clouds.   Here we consider velocities obtained with {\it
Gaia} DR2 proper motions \citep{brown2018,
lindegren2018,katz2018}. Note that the sky-on velocities of
OB-associations derived from {\it Gaia} DR1 and from  {\it Gaia}
DR2 proper motions differ on average by  2 km s$^{-1}$ \citep[for
more details see][]{melnik2017,melnik2018}.

For comparison with models we use  the residual velocities of
OB-associations which characterize non-circular motions in the
Galactic disc. The residual velocities are determined as
differences between the observed heliocentric velocities and the
velocities due to the Galactic circular rotation curve and the
solar motion towards the apex ($V_{res}=V_{obs}-V_{rot}-V_{ap}$).
The radial and azimuthal components, $V_R$ and $V_T$, of the
residual velocity are positive if they are directed away from the
Galactic center and  in the sense of Galactic rotation,
respectively. The residual velocity along z-axis, $V_z$, is
positive in the direction toward the North Galactic pole. The
parameters of the rotation curve and solar motion are derived from
the entire sample of OB-associations with known line-of-sight
velocities and Gaia DR2 proper motions
\citep{melnik2017,melnik2018}.  Residual velocities determined
with respect to a self-consistent rotation curve are practically
independent of the choice of the value of $R_0$.

\begin{figure}
\resizebox{\hsize}{!}{\includegraphics{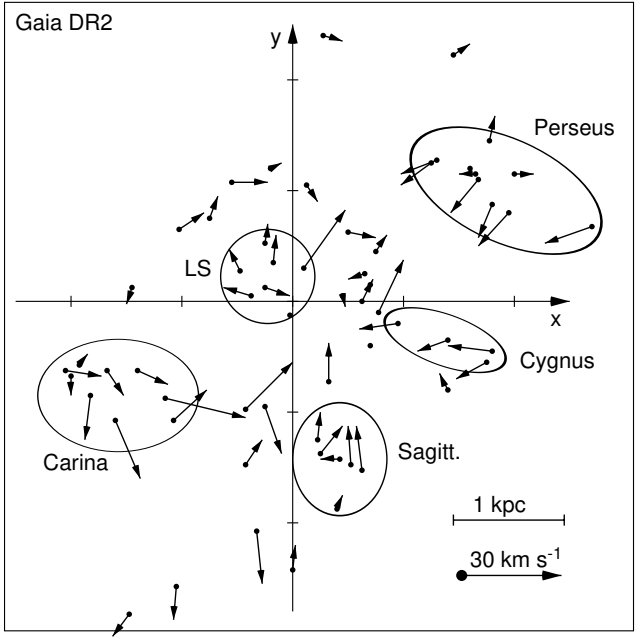}}
\caption{Distribution of the residual velocities of
OB-associations in the Galactic plane. The residual velocities are
derived with the use of  Gaia DR2 proper motions. OB-associations
with residual velocities $|V_R|$ and $|V_T|$ smaller than 3 km
s$^{-1}$ are shown as the black circles without any vector. The
ellipses indicate the positions of the Sagittarius, Carina,
Cygnus, Local System (LS), and Perseus stellar-gas complexes. The
x- and y- axes are directed towards the Galactic rotation and
away from the Galactic center, respectively.  The Sun is at the
origin.} \label{res_vel}
\end{figure}

Figure~\ref{res_vel} shows the residual velocities of
OB-associations in the Galactic plane. To mitigate the random
errors, we average the residual velocities of OB-associations
within the volumes of the stellar-gas complexes identified by
\citet{efremov1988}. Table 1 gives the name of the stellar-gas
complex, its Galactocentric distance $R$, the list of
OB-associations related to it, the range of their Galactic
longitudes $l$ and  heliocentric distances $r$, their average
residual velocities: $V_R$, $V_T$ and $V_z$. The OB-associations
having at leat 2 stars with known line-of-sight velocities and
Gaia DR2 proper motions are considered.

The Galactocentric distance of the Sun is adopted to be $R_0=7.5$
kpc \citep{glushkova1998, nikiforov2004, feast2008,
groenewegen2008, reid2009b, dambis2013, francis2014, boehle2016,
branham2017}. The choice of $R_0$ in the range 7--9 kpc has small
influence on the residual velocities.

Figure~\ref{res_vel}  and Table 1 indicate  that the majority of
OB-associations in the Perseus  complex have the radial component
of the residual velocity, $V_R$, directed toward the Galactic
center while the  velocities $V_R$ of most of OB-associations in
the Sagittarius and Local System complexes are directed away from
the Galactic center. As for the azimuthal  residual velocities,
the majority of OB-associations in the Perseus complex have $V_T$
directed in the sense opposite that of Galactic rotation while
$V_T$ is close to zero in the Sagittarius and Local System
complexes. It is just the residual velocities in the Sagittarius,
Perseus and Local System stellar-gas complexes that   can be
reproduced in present dynamical models.

However, the  residual velocities in the Cygnus and Carina
complexes are still remaining a stumbling block for numerical
simulations. That concerns both types of models: those with
analytical bars and N-body simulations
\citep{melnikrautiainen2009, rautiainen2010}.

Table~1 also shows  that the average  velocities in the
z-direction, $V_z$, are  close to zero. Here we suppose that
motions in the Galactic plane and in the z-direction are
independent that allows us to use 2D models.

Note that models considered must also reproduce the Galactic
rotation curve determined for the sample of OB-associations. To
avoid the systematical effects, we must  use the same sample of
objects for a study of residual velocities and for a determination
of the parameters of rotation curve. The rotation curve derived
from the velocities of OB-associations is nearly flat and
corresponds to the angular velocity at the solar distance of
$\Omega_0=31\pm1$ km s$^{-1}$ kpc$^{-1}$
\citep{melnik2017,melnik2018}.

\begin{table*}
  \caption{ The  observed residual velocities of OB-associations in the stellar-gas complexes with {\it Gaia} DR2 data}
  \begin{tabular}{lccccccl}
  \hline
  Complex& {\it R} & $V_R$ & $V_T$ & $V_z$ & {\it l} & {\it r} & Associations \\
    &  kpc & km s$^{-1}$ & km s$^{-1}$ & km s$^{-1}$   & deg. & kpc &  \\
  \\[-7pt]\hline\\[-7pt]
  Sagittarius & 6.0 & $+7.5\pm2.1$ & $-0.3\pm1.7$ & $-0.6\pm1.8$  & 8--23$^\circ$  & 1.3--1.9 & Sgr OB1,  OB4, OB7, Ser OB1, OB2, Sct OB3;\\
  Carina & 6.9 & $-6.2\pm2.6$ & $+6.2\pm2.8$ & $-1.9\pm0.7$ &  286--315$^\circ$  & 1.5--2.1 & Car OB1, OB2, Cru OB1, Cen OB1,\\
  & & & & &   &  & Coll 228, Tr 16, Hogg 16, NGC 3766, 5606;\\
  Cygnus & 7.3 & $-4.3\pm1.3$ & $-10.3\pm1.4$  & $+2.0\pm1.4$ & 73--78$^\circ$  & 1.0--1.8 & Cyg OB1, OB3, OB8, OB9; \\
  Local System & 7.8 & $+5.4\pm2.6$ & $+1.2\pm2.6$ & $-0.1\pm0.5$  & 0--360$^\circ$  & 0.3--0.6 & Per OB2, Ori OB1, Mon OB1, Vela OB2, \\
  & & & &  & & & Coll 121, 140; \\
  Perseus & 8.8 & $-4.7\pm2.2$ & $-4.4\pm1.7$  &  $+0.5\pm0.6$ & 104--135$^\circ$  & 1.8--2.8 & Cep OB1, Per OB1,  Cas OB1, OB2, OB4, \\
  & & & & & & & OB5, OB6, OB7, OB8, NGC 457;\\
  \hline
\end{tabular}
\end{table*}

\section{Models}

I have built several models with analytical Ferrers bars
\citep{ferrers1877} which  can reproduce the kinematics in the
Perseus, Sagittarius, and Local System stellar-gas complexes. Some
of them are discussed here.

Table 2 lists the general parameters of model 1: the time of
simulation $T$, the time step of integration $\Delta t$, the
number of  particles $N$.  We neglect the self-gravity between
model particles. The massless test particles can be thought as
low-mass gas clouds  moving in the potential created by the
stellar subsystem. The orbits of model particles are calculated
with the use of leapfrog method.

All models include a bar,  a disc, a bulge and halo, whose
parameters are given in Table 2. The bar is modelled as a Ferrers
ellipsoid with volume-density distribution $\rho$ defined as
follows:

\begin{equation}
\rho = \left\{ \begin{array}{l}
\rho_0 (1-\mu^2)^n, \ \mu \leq 1\\
\\
0, \ \mu > 1,\\
\end{array} \right.
\label{kalnajs}
\end{equation}

\noindent {\bf where $\mu$ equals $\mu^2=x^2/a^2+(y^2 + z^2)/b^2$}
but $a$ and $b$ are the lengths of the major and minor semi-axes
of the bar, respectively.  Here we consider 2D-models, so $z=0$.
The acceleration created by the bar depends on the mass of the bar
$M_b$, semi-axes $a$ and $b$, the  coordinates ($x$, $y$) reckoned
with respect to the bar axes and the power index $n$
\citep{freeman1972, pfenniger1984, binney1987,sellwood1993}.

The angular velocity of the bar, $\Omega_b$, providing the best
agrement with observations appears  to be $\Omega_b=50$ km
s$^{-1}$ kpc$^{-1}$. The non-axisymmetric perturbations of the bar
grows slowly approaching the full strength during $T_{gr}=492$ Myr
equal to four bar revolution periods. However, $m=0$ component of
the bar is included in  models from the beginning.  It can be
interpreted as a pre-existent disc-like bulge
\citep{athanassoula2005}. The mass of the bar, $M_b=1.30\times
10^{10}$M$_\odot$, agrees with  other estimates
\citep[e.g.][]{dwek1995}.

Models include an exponential disc with the scale length $R_d$:

\begin{equation}
\Sigma = \Sigma_0\;e^{-R/R_d},
 \label{edisk}
\end{equation}

\noindent where $\Sigma$ and $\Sigma_0$ are the surface densities
at the radius $R$ and at the galactic center, respectively. The
velocity of the rotation curve, $V_c$, produced by exponential
disc is determined by the following relation:

\begin{equation}
V_c^2 = 4\pi G \Sigma_0 R_d y^2 [I_0(y)K_0(y)-I_1(y)K_1(y)],
 \label{freeman}
\end{equation}

\noindent where $y=0.5R/R_d$  while  $I_n$ and $K_n$   are
modified Bessel functions of order $n$ of the first and second
kinds, respectively \citep{freeman1970, binney1987}.

Here the mass of the disc is adopted to be $M_d=3.5\times
10^{10}$M$_\odot$. To compare with a disc mass in N-body
simulations we should sum the mass of the disc and some part of
that of the bar. The total value, $3.5\textrm{--}4.8\times
10^{10}$M$_\odot$, is consistent with other estimates of the
Galactic disc mass, 3.5--5.0$\times 10^{10}$M$_\odot$
\citep{shen2010, fujii2019}.

The classical bulge determines the  potential in the galactic
center, it is modelled by a Plummer sphere \citep[for
example,][]{binney1987} whose rotation curve is defined by
following expression:

\begin{equation}
 V_c^2(R)=\frac{GM_{bg}{R}^2}{(R^2+R_{bg}^2)^{3/2}},
\end{equation}

\noindent where $M_{bg}$ and $R_{bg}$ are  the mass and
characteristic length of the bulge. The mass of the Galactic
classical bulge is expected to lie in the range 3--6 $\times
10^9$M$_\odot$ and the adopted value is $M_{bg}=5\times 10^{10}$
here \citep[e.g.,][]{dehnen1998,nataf2017,fujii2019}.

The halo dominates on the galactic periphery. It is modelled as an
isothermal sphere with following rotation curve:

\begin{equation}
 V_c^2(r)= V_{max}^2 \frac{R^2}{R^2+R_h^2},
\end{equation}

\noindent where $V_{max}$ is the asymptotic maximum on the halo
rotation curve and $R_h$ is the core radius.

\begin{figure*}
\resizebox{\hsize}{!}{\includegraphics{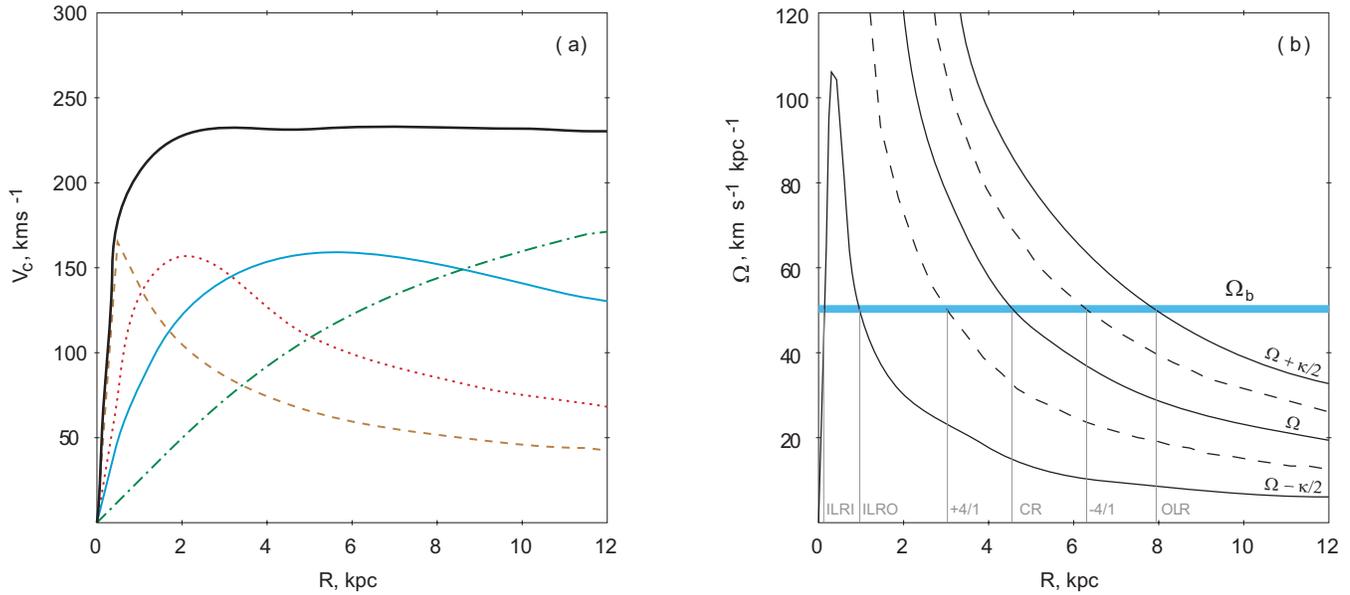}} \caption{(a)
Model rotation curves. The thick black line shows the total
rotation curve while the dashed, dotted, solid and dash-dotted
lines (colored brown, red, blue,  green in electronic edition)
indicate the contribution of the bulge, bar, disc and halo,
respectively. (b) Dependence of the angular velocities on the
galactocentric distance $R$. The continuous curves represent the
angular velocities $\Omega$ and $\Omega \pm \kappa/2$ while the
dashed lines indicate $\Omega \pm \kappa/4$. The horizontal thick
line (colored blue in electronic edition) shows the angular
velocity of the bar, $\Omega_b=50$ km s$^{-1}$ kpc$^{-1}$. The
resonance distances are determined by its intersections with the
curves of angular velocities.} \label{rot_curve}
\end{figure*}

Figure~\ref{rot_curve}(a) shows the total rotation curve produced
by the gravitation of the bulge, bar, disc and halo. The total
rotation curve is nearly flat with the average azimuthal velocity
of $\Theta=232$ km s$^{-1}$. This model value corresponds to the
angular velocity at the solar distance ($R_0=7.5$ kpc) of
$\Omega_0=30.9$ km s$^{-1}$ kpc$^{-1}$ and is  consistent with
observations, $\Omega_0=31\pm1$ km s$^{-1}$
kpc$^{-1}$\citep{melnik2017,melnik2018}. Model and observed
rotation curves are in good agreement with each other, at least in
the 3-kpc solar neighborhood.

Figure~\ref{rot_curve}(b) demonstrates the positions of the
resonances which correspond to the intersections of the horizontal
line indicating the angular velocity of the bar ($\Omega_b=50$ km
s$^{-1}$kpc$^{-1}$) with the appropriate curves of  angular
velocities. Note that present models include two inner Lindblad
resonances (ILRs): outer (ILRO) and inner (ILRI) ones. The CR of
the bar lies at the radius of 4.6 kpc near the end of the Ferrers
ellipsoid ($a=4.2$ kpc), so the model bar   is dynamically fast
\citep{debattista2000, rautiainen2008}.

The models can be rescaled for  a bit different values of the
solar Galactocentric distance  $R_0$. It is possible due to the
fact that the Galactic rotation curve is flat in the solar
neighborhood. If the ratio of the new and old values of $R_0$ is
$q=(R_{new}/R_{old})$ then the masses of the bulge, bar and disk
must be changed by a factor of $\sim q^2$ but the asymptotic
velocity of the halo -- by a factor of $\sim q$. New rotation
curve will be also flat in the solar neighborhood.

If test particles approach  each other at a small distance,
$\varepsilon$, they can collide with each other inelastically
imitating the behavior of a gas subsystem \citep{brahichenon1977,
levinson1981, roberts1984}. Collisions are adopted to be
absolutely inelastic, so the velocities of two gas particles after
a collision are the same $\bf v_1'= v_2'$ equal to $({\bf
v_1}+{\bf v_2})/2$, where $\bf v$ and $\bf v'$ are the velocities
before and after a collision, respectively.

The changes of velocities due to collisions at each step are
performed before the leapfrog integration. To accelerate the
computation of collisions  we sort particles  in the ascending
order of the coordinate $x$  at each step of integration as it is
supposed by \citet{salo1991}.

There is a danger that a pair of gas particles can fall into
"permanent collisions" \citep{brahichenon1977}, i. e. the same two
particles would collide at each step of integration. The galactic
differential rotation can tear some close pairs of colliding
particles  but it is powerless for particles lying at the same
galactic radius. The main remedy against "permanent collisions" is
quite large  velocity dispersion in radial direction maintained
above some minimal level, $\sigma_{min}$, throughout the galactic
disc. "Permanent collisions" can be avoided if the average
distance passed by a particle relative to another during one step
of integration $\Delta t$ is always larger than the length of
collision, $\varepsilon$:

\begin{equation}
 \sqrt 2 \; \sigma_{min} \: \Delta t \ge \varepsilon.
\label{per_col}
\end{equation}

\noindent Epicyclic motions and perturbations from the bar
increase the velocity dispersion while collisions  decrease it.
The parameter $\varepsilon$ regulates the frequency of collisions
which grows with increasing $\varepsilon$. The choice of the
initial velocity dispersion $\sigma_0$ equal to 5 km s$^{-1}$ and
$\varepsilon$ of 0.05 pc (Table 2) provides the velocity
dispersion, $\sigma_R$, never dropping below 5 km s$^{-1}$, so the
condition~(\ref{per_col}) is always fulfilled  (see also section
4.4).

Two colliding gas particles have a probability, $P_c$, that one of
them  forms an OB-association which wouldn't  take part in
collisions during some time interval, $t_{ob}$. OB-particles move
ballistically but after time $t_{ob}$ they  transform  back into
gas particles resuming their ability to collide
\citep{roberts1984, salo1991}.  The interest to OB-particles is
due to the fact that they indicate the places with the highest
density of model particles and outline the positions of different
morphological structures. Generally, OB-particles only roughly
imitate the process of the formation of OB-associations. Values of
$P_{ob}$ and $t_{ob}$ are usually taken to be 10 per cent and 4
Myr, respectively (Table 2).

The initial surface density  of model particles in model 1 is
uniform  within the radius $R<11$ kpc (Table 2).

Model 1 is a basic model of our study.  But we also consider three
other models which differ from model 1 in one of the following
features: the initial distribution,  presence/absence of
collisions and the power index $n$ of the  density distribution
inside the Ferrers ellipsoid. Model 2 starts from an exponential
distribution of gas particles in the galactic plane with the scale
length of $r_d=2.5$ kpc but values of all other parameters
coincide with those of model 1. Model 3 is collisionless and that
is its only difference from model 1. Model 4 includes the Ferrers
bar whose density distribution is determined by the power index
$n=1$, i. e. its bar is less centrally concentrated than  in other
models, but the mass of the bar, $M_b$, and all other parameters
are the same as in model 1. Table 3 briefly characterizes models
1--4 and presents the total number of collisions, $N_c$, occurred
during the simulation time.

All models considered have the same angular velocity of the bar,
$\Omega_b=50$ km s$^{-1}$ kpc$^{-1}$. As the distribution of the
potential is the same in Models 1--3,  the locations of the
resonances must also be the same there. Formally, model 4 differs
from models 1--3 in the potential distribution but that weakly
affects the positions of the resonances. Table 4 presents the
resonance locations in models 1--3 and in model 4. It is seen that
the difference in the resonance radii doesn't exceed  0.1 kpc and
mainly concerns the ILRO and +4/1 resonance, but the radius of the
OLR is the same in all models considered.

The amount of non-axisymmetric perturbations produced by the bar
is usually estimated through the parameter $Q_T(R)$, which  is the
ratio of the maximal tangential force at some radius to the
azimuthally average radial force at the same radius:

\begin{equation}
 Q_T(R)= \frac{\max(|F_T|)}{<|F_R|>}.
\label{def_qt}
\end{equation}

\noindent The value of $Q_T$ varies with radius. Its  maximal
value  is named $Q_b$ and is usually used as a measure of the
strength of the bar:

\begin{equation}
 Q_b= \max[Q_T (R)]
\label{def_qb}
\end{equation}

\noindent \citep{sanders1980,combes1981,athanassoula1983}.

\begin{figure}
\resizebox{\hsize}{!}{\includegraphics{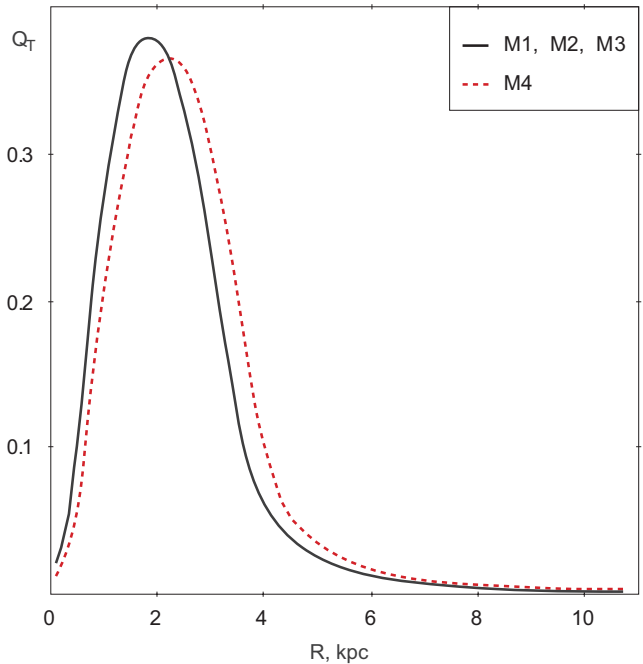}} \caption{
Variations of the parameter $Q_T$ along the galactic radius, $R$,
in models 1--3 and in model 4. The power index $n$ equals $n=2$
and $n=1$ in models 1--3 and  in model 4, respectively, while
other parameters determined the potential are the same. Maximal
value of $Q_T$ amounts to 0.380 in models 1--3 and to 0.367 in
model 4. However, at the distance of the OLR, 7.9 kpc, the value
of $Q_T$ is larger in model 4 ($Q_T=0.0074$) than in models 1--3
($Q_T=0.0057$).} \label{qt}
\end{figure}

Figure~\ref{qt} shows the  variations of $Q_T$ along the galactic
radius calculated for models 1--3 and for model 4.   Maximal
values of $Q_T$  are  0.380 (models 1--3) and 0.367 (model 4).
They are achieved at the distances of 1.8 and 2.2 kpc,
respectively. The value of the bar strength  $Q_b \approx 0.38$ is
quite expectable for galaxies with strong bars \citep{block2001,
buta2004, diaz-garcia2016}.  Note that at the distance of the OLR,
$R=7.9$ kpc, the value of $Q_T$ is larger in model 4
($Q_T=0.0074$) than in models 1--3 ($Q_T=0.0057$) by ~25 per cent.
This small preponderance of model 4  being amplified by the
resonance results in larger velocity perturbations produced by
model 4 in the solar neighborhood.

The value of the bar strength, $Q_b$, is sensitive to the choice
of the bulge mass: the larger $M_{bg}$ the smaller $Q_T$. For
example, the increase of $M_{bg}$ from 5 to $9\times10^9$M$_\odot$
results in the decrease of $Q_b$ from 0.38 to 0.34.

\begin{figure*}
\resizebox{17 cm}{!}{\includegraphics{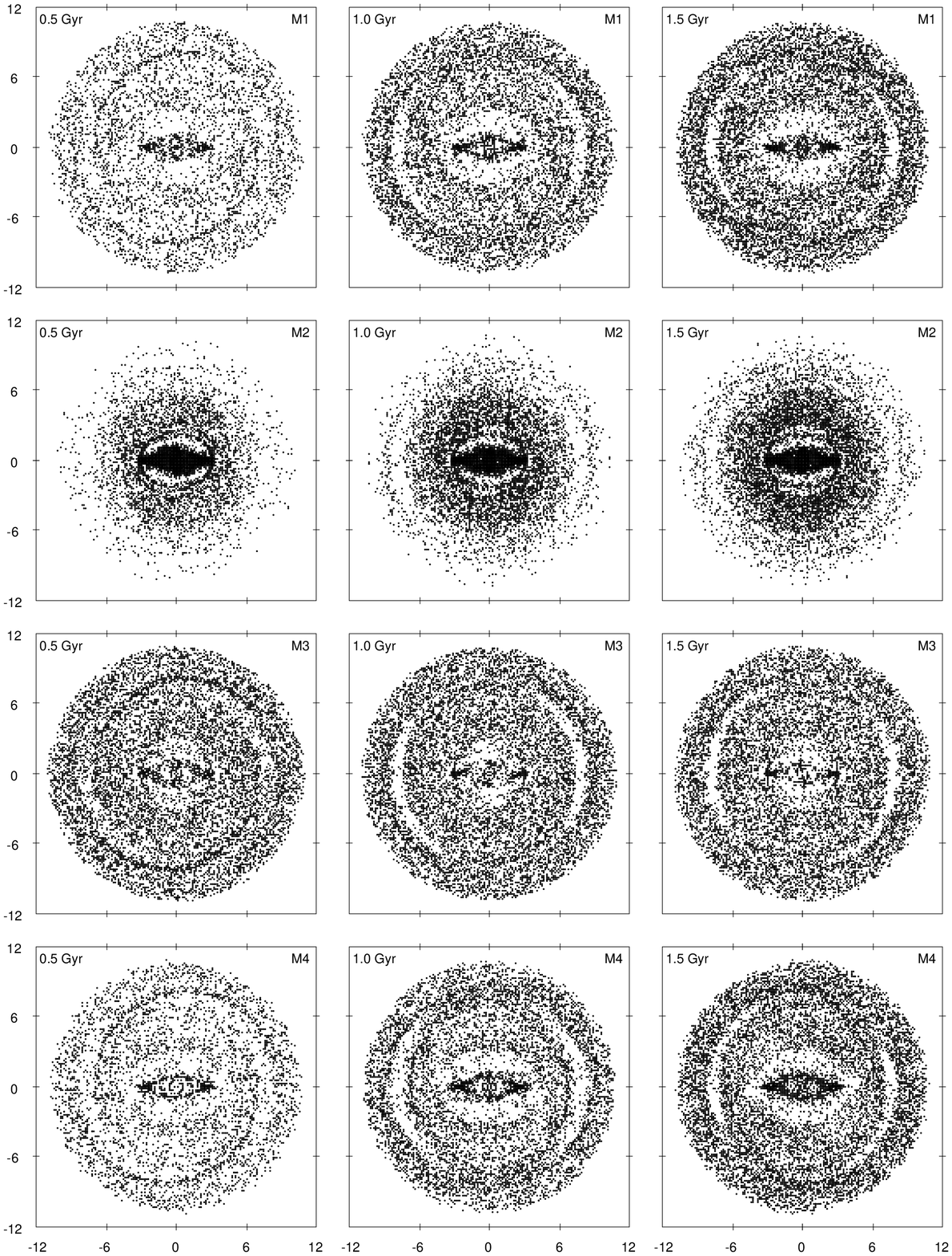}} \caption{
Distribution of model particles at three time moments: 0.5, 1.0
and 1.5 Gyr. Frames related to models 1, 2 and 4  display the
distribution of  OB-particles while model 3 demonstrates the
distribution of 10 per cent of collisionless particles. The size
of frames is $24 \times 24$ kpc.} \label{distrib}
\end{figure*}

Figure~\ref{distrib} shows the distribution of model particles at
three time moments: 0.5, 1.0 and 1.5 Gyr. The frames related to
models 1, 2 and 4 demonstrate the distribution of OB-particles.
Model 3 is collisionless, so OB-particles aren't forming there,
the corresponding frames merely present the distribution of 10 per
cent of model particles. We can see that model discs form nuclear
rings ($\sim 0.5$ kpc) and conspicuous outer rings $R_2$ ($\sim
9.0$ kpc). Model galaxies also produce outer rings $R_1$ ($\sim 6$
kpc) and inner rings ($\sim 3$ kpc) which are mainly noticeable in
the density profiles (section 4.3). Note that all models
demonstrate the diamond-shape structures which are located inside
the Ferrers ellipsoids and  indicate the places of most densely
populated bar orbits. These structures aren't inner rings which
usually have more round shapes and are forming outside the bar.
Model 2 ($t=0.5$ Gyr) gives a good example of an inner ring which
touches the bar only at the bar ends. Models 1 and 3 also include
inner rings but they are hardly visible in Figure~\ref{distrib}
(see section 4.3). The nuclear and inner rings are forming quickly
and are already existent at the time $t=0.5$ Gyr when the bar
acquires its full strength. The outer rings are growing slower:
they appear as pseudorings at $t=0.5$ Gyr and take pure elliptical
shape at the time $t\approx 1.0$ Gyr. Once formed, the outer rings
exist to the end of simulation.

\begin{table}  \caption{General parameters of model 1}
  \begin{tabular}{ll}
  \hline
Simulation time &  $T=2$  Gyr  \\
Step of integration     &  $\Delta t=0.01$ Myr \\
Number of particles     &  $N=10^5$ \\
\hline
Bulge & $R_{bg}=0.30$  kpc    \\
      & $M_{bg}=5$ 10$^{9}$M$_\odot$  \\
\hline
Bar   & $a=4.2$  and $b=1.35$ kpc \\
      & $M_b=1.30$ 10$^{10}$M$_\odot$ \\
      & $\Omega_b=50.0$ km s$^{-1}$ kpc$^{-1}$  \\
      & $T_{gr}=492$ Myr  \\
\hline
Disc  & exponential, $R_d=2.5$ \\
      & $M_d=3.5$ 10$^{10}$M$_\odot$ \\
\hline
Halo &  $R_h=8$  kpc   \\
     &  $V_{max}=206$  km s$^{-1}$  \\
\hline
Collisions &  absolutely inelastic   \\
           &  $\varepsilon=0.05$  pc   \\
\hline OB-particles &  $t_{ob}=4$  Myr -- lifetime   \\
 &  $P_c=0.1$  -- probability   \\
\hline
Initial distribution & uniform  within $R<11$ kpc \\
                      &   $\sigma_0=5$  km s$^{-1}$ \\
\hline
\end{tabular}
\label{general}
\end{table}

\begin{table}
  \caption{Characteristics of models 1--4}
  \begin{tabular}{lccc}
   \hline
  Model  & Initial distribution    & $n$ & $N_c$   \\
  \\[-7pt]\hline\\[-7pt]
   1  & uniform  within $R<11$ kpc   & $n=2$ & 3.6 10$^7$\\
   2  & exponential, $r_d=2.5$ kpc   & $n=2$ & 4.7 10$^7$\\
   3  & uniform  within $R<11$ kpc   & $n=2$ & 0          \\
   4  & uniform  within $R<11$ kpc   & $n=1$ & 3.6 10$^7$ \\
\hline
\end{tabular}
\end{table}

\begin{table}
  \caption{ Locations of the resonances}
  \begin{tabular}{lccc}
   \hline
  Name & Definition & Models 1--3 & Model 4  \\
       &            & $R$, kpc & $R$, kpc  \\
  \\[-7pt]\hline\\[-7pt]
   OLR  &  $\kappa/(\Omega-\Omega_b)=-2/1$ & 7.91  & 7.91\\
   -4/1 &  $\kappa/(\Omega-\Omega_b)=-4/1$ & 6.29  & 6.29\\
   CR   &  $\Omega=\Omega_b$               & 4.61  & 4.62\\
   +4/1 &  $\kappa/(\Omega-\Omega_b)=4/1$  & 3.01  & 2.92\\
   ILRO &  $\kappa/(\Omega-\Omega_b)=2/1$  & 0.97  & 0.92\\
   ILRI &  $\kappa/(\Omega-\Omega_b)=2/1$  & 0.13  & 0.13\\
\hline
\end{tabular}
\end{table}

\section{Results}

\subsection{Kinematics of model particles in the solar
neighborhood}

The epicyclic motions of model particles located near the OLR of
the bar are adjusted   in accordance with the  perturbations
coming from the bar that results in the formation of conspicuous
systematic velocities.

\begin{figure}
\resizebox{\hsize}{!}{\includegraphics{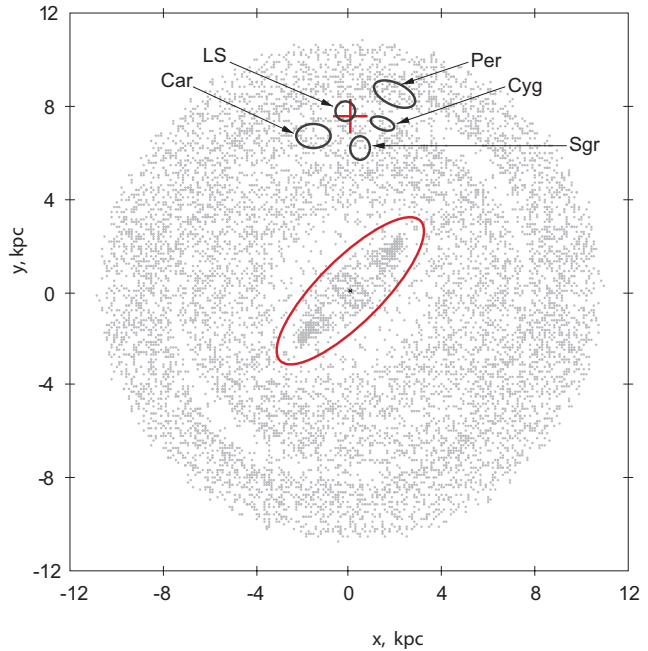}}
\caption{Distribution of OB particles (gray points) in the
galactic plane. Also shown are the boundaries of the stellar-gas
complexes: Sagittarius, Carina, Cygnus, Local System (LS) and
Perseus. Model 1 at $t=1.5$ Gyr is considered. The position of the
Sun is indicated by a large cross. The location of the Ferrers
ellipsoid is also marked. The position angle of the Sun with
respect to the bar is supposed to be $\theta_b=45^\circ$.  As our
Galaxy is traditionally considered rotating clockwise (i. e. as if
being observed from the  North Galactic Pole), the model galaxy is
also turned to rotate clockwise. The Sagittarius, Carina, Cygnus
and LS complexes lie in the vicinity of the outer ring $R_1$ but
the Perseus complex is related to the outer ring $R_2$. }
\label{m1_d}
\end{figure}

Figure~\ref{m1_d} shows the distribution of OB particles in the
galactic plane in model 1 at $t=1.5$ Gyr and the boundaries of the
Sagittarius, Carina, Cygnus, Local System (LS) and Perseus
stellar-gas complexes. Model velocities in the stellar-gas
complexes are calculated as average velocities of model particles
(gas+OB) located inside the boundaries of the complexes at a
moment considered. The model residual velocities are determined
with respect to the model rotation curve. The velocities in the
complexes are derived every 10 Myr.  Note that at every moment
considered, the boundaries of  complexes include different sets of
model particles, but the position of the boundaries with respect
to the bar major axis remains the same corresponding to  the solar
position angle of $\theta_b=45^\circ$ (see section 4.2).

\begin{figure*}
\resizebox{ 17 cm}{!}{\includegraphics{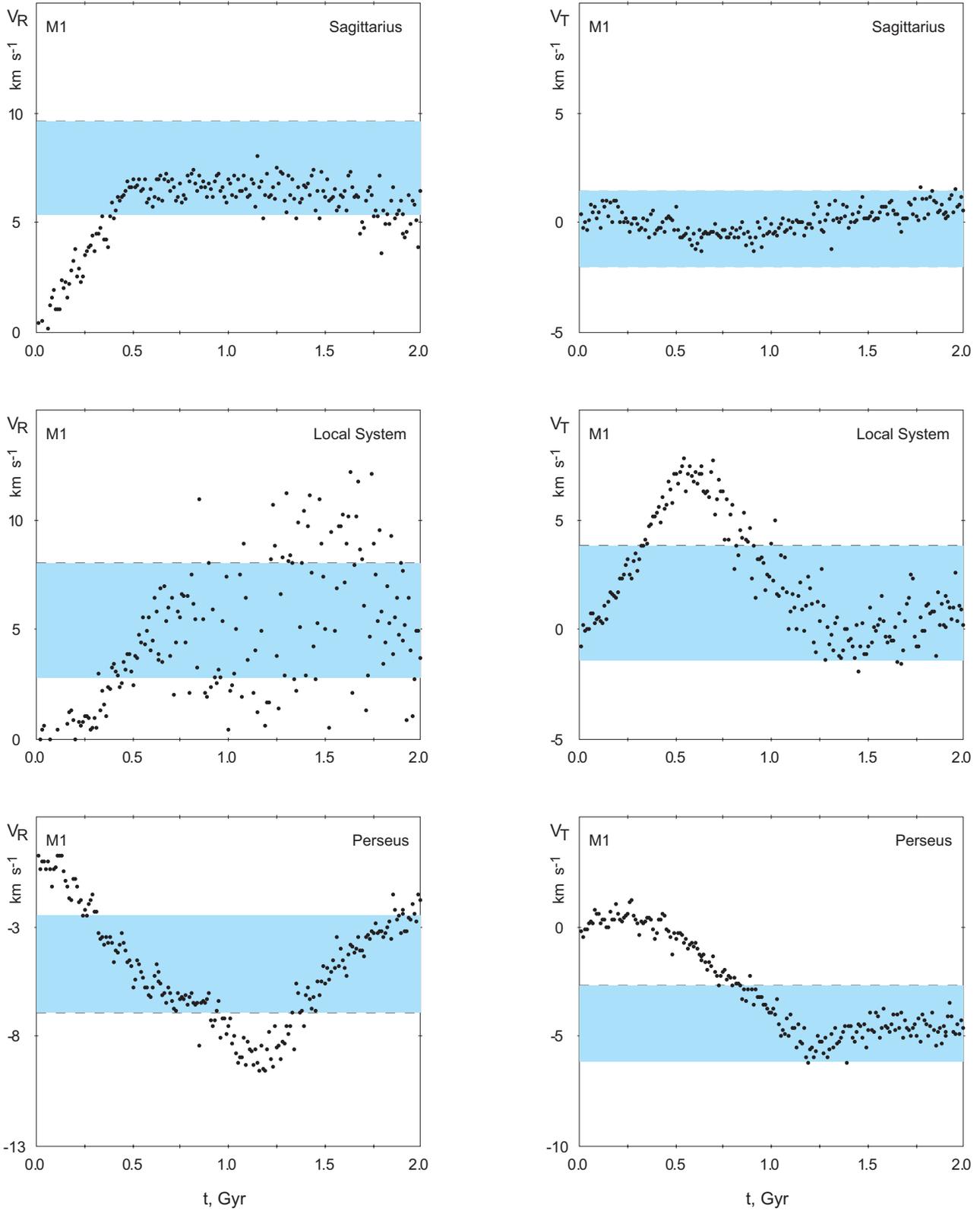}}
\caption{Comparison of model  and observed  residual velocities
calculated for the three stellar-gas complexes: Sagittarius, Local
System and Perseus. The left and right panels show radial and
azimuthal, $V_R$ and $V_T$, residual velocities, respectively. The
gray strips (colored blue in electronic edition) display the
uncertainties in determination of observed residual velocities,
$V_{obs}\pm\varepsilon_v$ (Table 1). Black circles indicate the
average velocities of model particles (gas+OB) located inside the
boundaries of stellar-gas complexes  in model 1 every 10 Myr. The
position angle of the Sun with respect to the bar is adopted to be
$\theta_b=45^\circ$.} \label{vel_comp}
\end{figure*}

Figure~\ref{vel_comp} demonstrates the residual velocities in the
Sagittarius, Local System and Perseus stellar-gas complexes
computed for model 1 at different time moments. We can see that
the scatter of model velocities is quite small (0.3--1.0 km
s$^{-1}$) everywhere except $V_R$-velocities in the Local System
where it  amounts to 3--4 km s$^{-1}$. The Local System is located
between two outer rings  and includes particles related to both of
them: $R_1$-objects have positive radial velocities while
$R_2$-objects have negative ones (see also Fig.~\ref{schema}).
Though the scatter of velocities $V_R$ in the Local System is
quite large, their average value is close to the observed one,
$V_R=5.4$ km s$^{-1}$ (Table 1).

Table 5 lists the average values of model residual velocities,
$V_R$ and $V_T$, in the three stellar-gas complexes computed for 7
time intervals: 0.6--0.8, 0.8--1.0, 1.0--1.2, 1.2--1.4, 1.4--1.6,
1.6--1.8, 1.8--2.0 Gyr, each of which includes 20 instantaneous
estimates. It also  presents the average time from the start of
simulations at the interval considered, $t$, and the average
number of model particles, $n$, which appear to be inside the
boundaries of the complexes. The last line indicates the observed
residual velocities in the corresponding complexes.

Figure~\ref{vel_comp} and Table 5 suggest  that there are a lot of
time moments when model and observed velocities in the three
stellar-gas complexes agree within the errors. The Sagittarius
complex demonstrates a good agreement between model and observed
velocities at the time interval 0.3--1.8 Gyr. In the Local System
the situation is more complicated: the model and observed radial
velocities, $V_R$, are consistent within the errors in 75 per cent
of time moments from the interval 0.5--2.0 Gyr but the azimuthal
velocities, $V_T$,  agree at the interval 0.8--2.0 Gyr. The
Perseus complex shows a good accordance between model and observed
radial velocities, $V_R$, at two time intervals, 0.5--1.0 and
1.4--1.8 Gyr, while the agreement in azimuthal velocities, $V_T$,
is reached at the time interval 1.0--2.0 Gyr. Hereafter,  $t=1.5$
Gyr will be regarded as a reference time moment  when model and
observed velocities in the Sagittarius, Local System and Perseus
stellar-gas complexes  are in good agreement.

\begin{figure*}
\resizebox{ 17 cm}{!}{\includegraphics{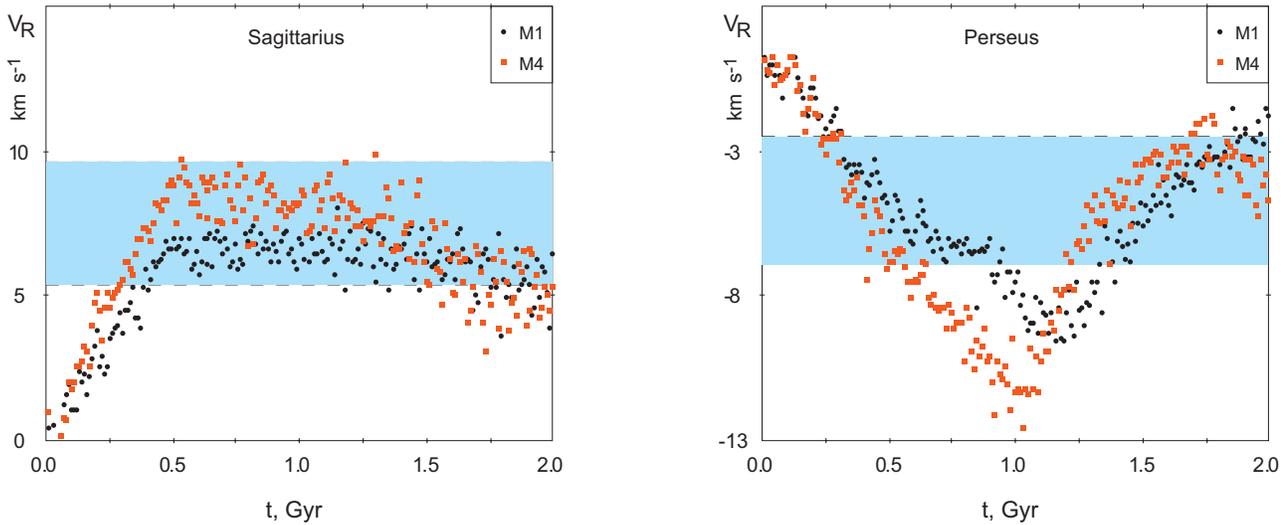}}
\caption{Variations of the radial residual velocity $V_R$  in the
Sagittarius and Perseus complexes  in model 1 and 4.  The absolute
values of the  velocities $V_R$ are larger in model 4 than in
model 1 what  can be connected with the larger value of $Q_T$ in
model 4 than in model 1 in the solar neighborhood. }
\label{vres_m1_m4}
\end{figure*}

Table 5 indicates that the residual velocities, $V_R$ and $V_T$,
produced by models 1--3 are nearly the same while  model 4 creates
a bit larger velocity perturbations. This difference  is
especially noticeable in the distribution of velocities $V_R$ in
the Sagittarius and Perseus complexes. To compare model 1 and 4 we
build Figure~\ref{vres_m1_m4} which shows the radial residual
velocities $V_R$ produced by  both models in the Sagittarius and
Perseus complexes. The absolute values of the radial velocities
$V_R$ are larger in model 4 than in model 1 what can be connected
with the larger value of $Q_T$  in model 4 than in model 1 in the
solar neighborhood (Fig.~\ref{qt}).

\begin{figure*}
\resizebox{\hsize}{!}{\includegraphics{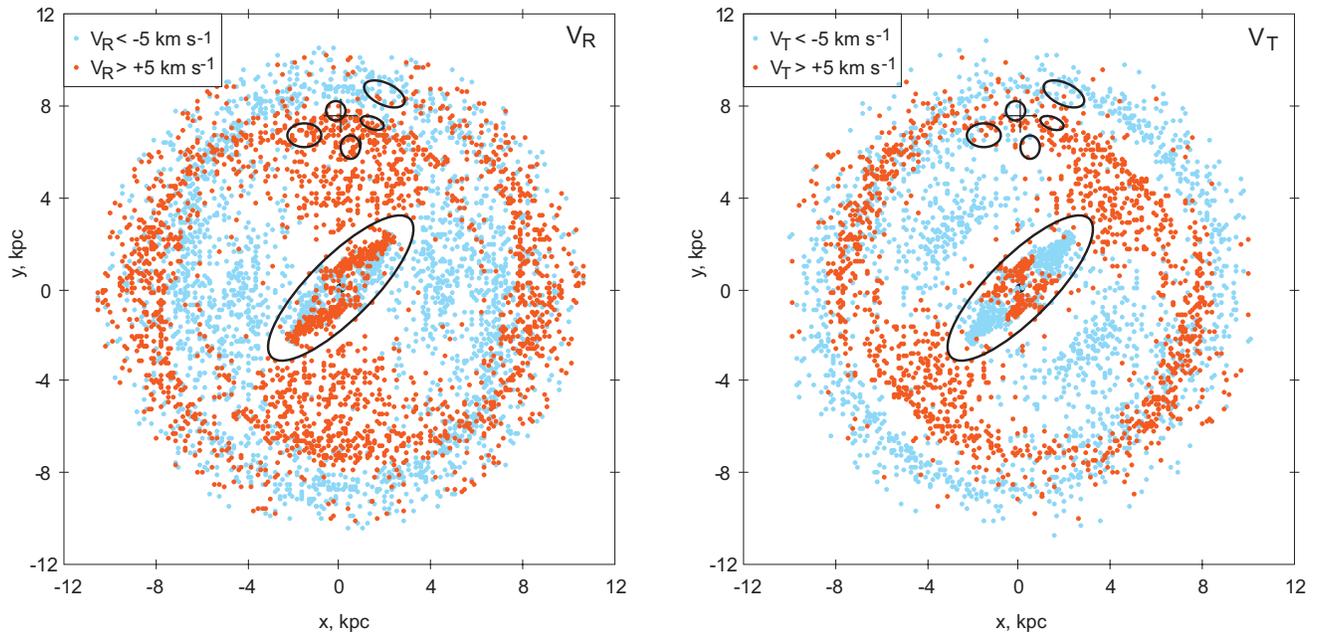}}
\caption{Distribution of  OB particles  with negative and positive
residual velocities in model 1 at $t=1.5$ Gyr.  The left and right
panels represent the distribution of  radial, $V_R$, and
azimuthal, $V_T$, residual velocities, respectively. Particles
with the conspicuous positive  velocities ($V_R>+5$  or $V_T>+5$
km s$^{-1}$) are indicated as dark-gray circles (colored red in
electronic edition) while those with the conspicuous negative
velocities ($V_R<-5$ or $V_T<-5$ km s$^{-1}$) are shown as
light-gray circles (colored blue in electronic edition). Particles
with the residual velocities close to zero ($|V_R|<5$ or $|V_T|<5
$ km s$^{-1}$) aren't shown here. It also represents the
boundaries of the Sagittarius, Carina, Cygnus, LS and Perseus
stellar-gas complexes. The model galaxy is turned to rotate
clockwise.  The location of the complexes is determined for the
solar position angle of $\theta_b=45^\circ$. The left panel
indicates that  the velocities $V_R$ of model particles are
positive in the Sagittarius, Carina, Cygnus and LS stellar-gas
complexes while they are negative in the Perseus complex.  The
right panel shows that   the velocities $V_T$ are negative in the
Perseus complex.} \label{neg_pos}
\end{figure*}

Figure~\ref{neg_pos} shows the distribution of OB particles with
negative ($V_{res}<5$ km s$^{-1}$) and positive ($V_{res}>5$ km
s$^{-1}$) residual velocities in model 1 at $t=1.5$ Gyr. Particles
with the residual velocities close to zero ($|V_{res}|<5$  km
s$^{-1}$) aren't shown there. The left panel indicates  that
velocities $V_R$ of model particles are positive in the
Sagittarius, Carina, Cygnus and Local System  stellar-gas
complexes while they are negative in the  Perseus complex.  The
right panel shows that   the velocities $V_T$ are negative in the
Perseus complex.

\begin{figure}
\resizebox{\hsize}{!}{\includegraphics{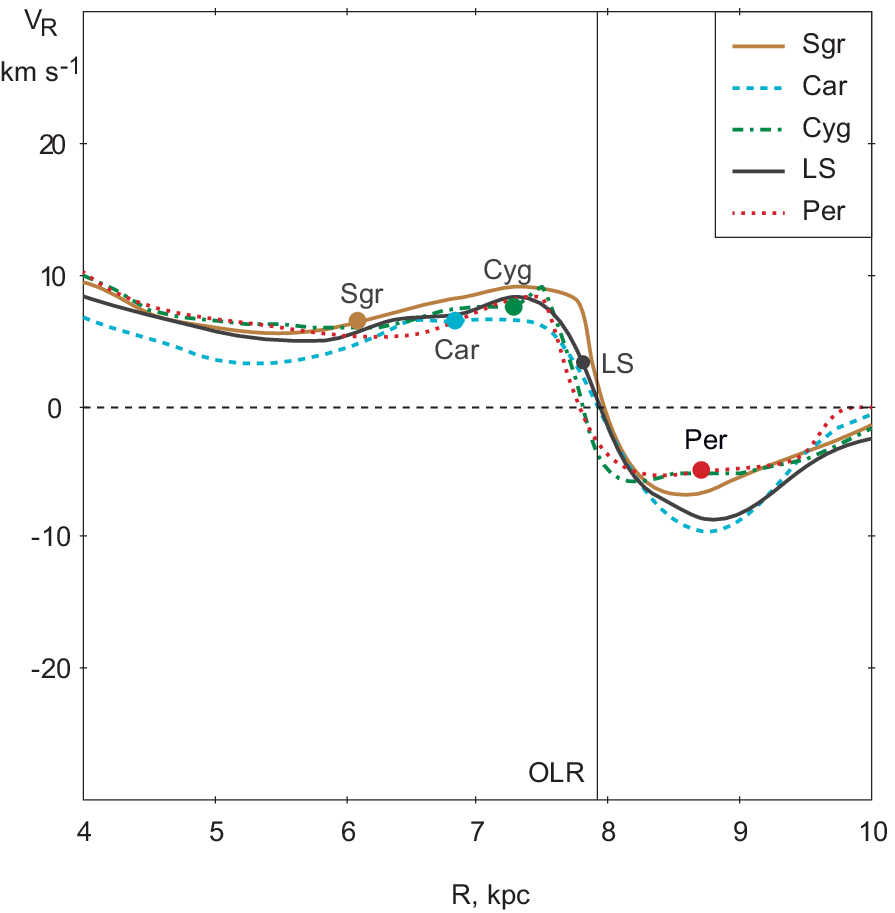}}
\caption{Variations of  the  radial  residual velocities, $V_R$,
along the Galactocentric distance, $R$, calculated for five
Galactic radius-vectors connecting the Galactic center with the
centers of the corresponding stellar-gas complexes: Sagittarius,
Carina, Cygnus, Local System and Perseus. The position angle of
the Sun with respect to the bar major axis is adopted to be
$\theta_b=45^\circ$. The vertical line indicates the radius of the
OLR. The model velocity $V_R$ at each point of the profiles is
computed as the average velocity of model particles (gas+OB)
located inside a small region with the  radius of  0.5-kpc and the
center lying on  the corresponding radius-vector. Model 1 at
$t=1.5$ Gyr is considered. All profiles demonstrate a sharp drop
in the velocity $V_R$ at the distance of the OLR.}
\label{vel_prof}
\end{figure}

Figure~\ref{vel_prof} demonstrates  the variations of the radial
velocity $V_R$ along the distance  $R$  calculated for five
radius-vectors connecting the Galactic center with the centers of
the corresponding stellar-gas complexes: Sagittarius, Carina,
Cygnus, Local System and Perseus.  These radius-vectors form a bit
different angles with the Sun-Galactic center line:
$\theta=4.0^\circ$ (Sgr), -13.3$^\circ$ (Car), +11.0$^\circ$
(Cyg), -1.6$^\circ$ (LS) and 12.8$^\circ$ (Per).  The velocity
$V_R$ at each point of the profiles is computed as the average
velocity of model particles (gas+OB) located inside a small circle
with the  radius of 0.5-kpc and  the center lying on the
corresponding radius-vector. Model 1 at $t=1.5$ Gyr is considered.
It is clear that all profiles demonstrate a sharp drop in the
velocity $V_R$ at the distance of the OLR. Generally, we can shift
the position of the OLR of the bar by choosing a different value
of $\Omega_b$, but if we like the velocity $V_R$ in the Local
System to be positive, then we would inevitably obtain positive
$V_R$ in the Sagittarius, Carina and Cygnus complexes which are
located at the smaller Galactocentric distances $R$ than that of
the Local System. However, the observed velocities $V_R$ in the
Carina and Cygnus complexes are negative (Table~1).

The uncertainty in a choice  of the value of the angular velocity
of the bar, $\Omega_b=50$ km s$^{-1}$ kpc$^{-1}$, is less than
$\pm2$ km s$^{-1}$ kpc$^{-1}$. If we choose $\Omega_b$ to be 52 km
s$^{-1}$ kpc$^{-1}$ than the radius of the OLR will be shifted by
0.3 kpc toward the Galactic center and the average velocities
$V_R$ in the Local System  will be negative. On the contrary, the
value of $\Omega_b=48$ km s$^{-1}$ kpc$^{-1}$ shifts  the OLR by
0.3 kpc away from the Galactic center and causes the velocities
$V_T$ in the Perseus region to be too small in absolute value,
$|V_T|<3$ km s$^{-1}$. All these changes cause a discrepancy with
observations.

Table 6 lists the average residual velocities of model particles,
$V_R$ and $V_T$, located within the boundaries of the Carina and
Cygnus stellar-gas complexes in model 1 at different time moments.
Other models give similar results. The bottom line indicates the
observed velocities.  It is clear that present models cannot
reproduce the observed velocities in the Carina and Cygnus
complexes.  Probably, some important physical processes which
determine the kinematics just in these two regions aren't included
into consideration.

\begin{table*}
\caption{Residual velocities, $V_R$ and $V_T$, calculated for the
Sagittarius, Local System and Perseus stellar-gas complexes in
models 1--4}
 \begin{tabular}{cc@{\qquad\qquad}rrr@{\qquad\qquad}rrr@{\qquad\qquad}rrr}
 \\[-7pt] \hline\\[-7pt]
 Complex&  & \multicolumn{3}{c}{Sagittarius$\qquad\qquad$} & \multicolumn{3}{c}{Local System$\qquad\qquad$} & \multicolumn{3}{c}{Perseus}  \\
  \\[-7pt] \hline\\[-7pt]
 Model& $t$ & $V_R\quad$ & $V_T\quad$ & $n\;$ & $V_R\quad\;$ & $V_T\quad\;$ & $n\;$ & $V_R\quad\;$ & $V_T\quad\;$ & $n\;$   \\
      & Gyr & km s$^{-1}$ & km s$^{-1}$ &  & km s$^{-1}$ & km s$^{-1}$ &  & km s$^{-1}$ & km s$^{-1}$ &   \\
\\[-7pt] \hline\\[-7pt]
Model 1 & 0.7 & $ 6.6\pm0.1$ & $-0.5\pm0.1$ & 178 & $ 5.1\pm0.3$ & $ 6.2\pm0.2$ & 186 & $-6.0\pm0.1$ & $-1.9\pm0.1$ & 378  \\
        & 0.9 & $ 6.7\pm0.1$ & $-0.5\pm0.1$ & 183 & $ 4.1\pm0.7$ & $ 3.4\pm0.2$ & 145 & $-6.9\pm0.2$ & $-3.1\pm0.1$ & 434  \\
        & 1.1 & $ 6.6\pm0.1$ & $-0.2\pm0.1$ & 182 & $ 2.5\pm0.7$ & $ 1.7\pm0.3$ & 103 & $-8.8\pm0.1$ & $-4.8\pm0.2$ & 488  \\
        & 1.3 & $ 6.5\pm0.1$ & $ 0.2\pm0.1$ & 184 & $ 6.5\pm0.7$ & $-0.0\pm0.2$ &  77 & $-7.7\pm0.2$ & $-5.3\pm0.1$ & 479  \\
        & 1.5 & $ 6.3\pm0.1$ & $ 0.4\pm0.1$ & 182 & $ 7.4\pm0.9$ & $-0.3\pm0.2$ &  74 & $-5.3\pm0.2$ & $-4.6\pm0.1$ & 457  \\
        & 1.7 & $ 5.9\pm0.2$ & $ 0.6\pm0.1$ & 184 & $ 7.4\pm0.7$ & $ 0.1\pm0.3$ &  79 & $-3.7\pm0.1$ & $-4.5\pm0.1$ & 452  \\
        & 1.9 & $ 5.4\pm0.2$ & $ 0.8\pm0.1$ & 191 & $ 5.0\pm0.5$ & $ 0.8\pm0.2$ &  94 & $-2.6\pm0.1$ & $-4.6\pm0.1$ & 449  \\
\\[-7pt] \hline\\[-7pt]
M2      & 0.7 & $ 6.9\pm0.1$ & $-0.6\pm0.1$ & 168 & $ 5.8\pm0.4$ & $ 5.5\pm0.2$ &  88 & $-6.6\pm0.2$ & $-2.2\pm0.1$ & 116  \\
        & 0.9 & $ 6.7\pm0.2$ & $-0.4\pm0.1$ & 165 & $ 4.9\pm0.9$ & $ 2.7\pm0.2$ &  69 & $-8.3\pm0.3$ & $-3.8\pm0.1$ & 143  \\
        & 1.1 & $ 6.3\pm0.1$ & $-0.3\pm0.1$ & 166 & $ 2.5\pm0.7$ & $ 0.9\pm0.3$ &  49 & $-10.3\pm0.2$ & $-5.8\pm0.2$ & 157  \\
        & 1.3 & $ 6.2\pm0.1$ & $-0.1\pm0.1$ & 173 & $ 7.4\pm1.4$ & $-1.0\pm0.3$ &  38 & $-9.7\pm0.3$ & $-6.7\pm0.2$ & 163  \\
        & 1.5 & $ 6.2\pm0.1$ & $ 0.2\pm0.1$ & 172 & $ 8.4\pm1.0$ & $-1.4\pm0.3$ &  34 & $-5.7\pm0.3$ & $-5.9\pm0.2$ & 147  \\
        & 1.7 & $ 5.8\pm0.2$ & $ 0.6\pm0.1$ & 170 & $ 7.9\pm0.8$ & $-1.0\pm0.4$ &  39 & $-3.8\pm0.2$ & $-6.1\pm0.2$ & 148  \\
        & 1.9 & $ 5.2\pm0.1$ & $ 0.6\pm0.1$ & 175 & $ 6.6\pm0.5$ & $-0.3\pm0.2$ &  46 & $-1.7\pm0.2$ & $-6.1\pm0.2$ & 150  \\
\\[-7pt] \hline\\[-7pt]
M3      & 0.7 & $ 6.6\pm0.1$ & $-0.3\pm0.1$ & 182 & $ 5.9\pm0.3$ & $ 5.8\pm0.3$ & 182 & $-5.8\pm0.1$ & $-1.8\pm0.1$ & 372  \\
        & 0.9 & $ 6.5\pm0.1$ & $-0.3\pm0.1$ & 184 & $ 4.4\pm0.8$ & $ 3.2\pm0.2$ & 144 & $-6.9\pm0.2$ & $-3.1\pm0.1$ & 429  \\
        & 1.1 & $ 6.7\pm0.1$ & $ 0.0\pm0.1$ & 180 & $ 2.9\pm0.6$ & $ 0.7\pm0.2$ & 104 & $-8.8\pm0.2$ & $-4.6\pm0.1$ & 475  \\
        & 1.3 & $ 6.4\pm0.1$ & $ 0.2\pm0.1$ & 185 & $ 6.0\pm0.8$ & $-0.6\pm0.3$ &  74 & $-8.0\pm0.2$ & $-5.2\pm0.1$ & 462  \\
        & 1.5 & $ 6.2\pm0.1$ & $ 0.7\pm0.1$ & 185 & $ 9.0\pm0.7$ & $-1.1\pm0.2$ &  71 & $-5.0\pm0.2$ & $-4.6\pm0.1$ & 439  \\
        & 1.7 & $ 5.8\pm0.2$ & $ 1.1\pm0.1$ & 190 & $ 7.4\pm0.4$ & $-0.4\pm0.2$ &  78 & $-3.4\pm0.1$ & $-4.6\pm0.1$ & 441  \\
        & 1.9 & $ 5.2\pm0.2$ & $ 1.2\pm0.1$ & 190 & $ 5.8\pm0.3$ & $ 0.1\pm0.2$ &  93 & $-2.0\pm0.1$ & $-4.8\pm0.1$ & 441  \\
\\[-7pt] \hline\\[-7pt]
M4      & 0.7 & $ 8.4\pm0.2$ & $-0.7\pm0.1$ & 174 & $ 6.6\pm0.5$ & $ 5.4\pm0.3$ & 171 & $-8.3\pm0.2$ & $-2.9\pm0.1$ & 385  \\
        & 0.9 & $ 8.2\pm0.1$ & $-0.1\pm0.1$ & 175 & $ 4.4\pm0.8$ & $ 2.7\pm0.3$ & 122 & $-10.3\pm0.2$ & $-4.9\pm0.2$ & 470  \\
        & 1.1 & $ 8.2\pm0.2$ & $ 0.6\pm0.1$ & 181 & $ 4.5\pm0.8$ & $ 0.2\pm0.3$ &  74 & $-9.8\pm0.3$ & $-5.8\pm0.1$ & 475  \\
        & 1.3 & $ 7.7\pm0.2$ & $ 1.1\pm0.1$ & 191 & $11.1\pm0.7$ & $-1.6\pm0.4$ &  60 & $-5.7\pm0.2$ & $-4.8\pm0.1$ & 427  \\
        & 1.5 & $ 6.6\pm0.2$ & $ 1.5\pm0.1$ & 194 & $ 8.6\pm0.7$ & $-0.6\pm0.2$ &  67 & $-3.9\pm0.2$ & $-4.9\pm0.1$ & 447  \\
        & 1.7 & $ 5.4\pm0.2$ & $ 1.3\pm0.2$ & 188 & $ 4.8\pm0.6$ & $ 1.3\pm0.2$ &  88 & $-2.9\pm0.2$ & $-5.1\pm0.1$ & 437  \\
        & 1.9 & $ 5.2\pm0.2$ & $ 0.7\pm0.1$ & 192 & $ 2.5\pm0.4$ & $ 2.2\pm0.2$ & 128 & $-3.7\pm0.2$ & $-4.0\pm0.1$ & 409  \\
\\[-7pt] \hline\\[-7pt]
\multicolumn{2}{l}{Observations} & $7.5\pm2.1$ & $-0.3\pm1.7$ & &  $5.4\pm2.6$ & $1.2\pm2.6$ && $-4.7\pm2.2$ & $-4.4\pm1.7$ & \\
    \\[-7pt] \hline\\[-7pt]
  \end{tabular}
 \label{tab5}
\end{table*}


\begin{table*}
\caption{Residual velocities, $V_R$ and $V_T$, computed for  the
Carina and Cygnus stellar-gas complexes in model 1}
 \begin{tabular}{c@{\qquad}rrr@{\qquad\qquad}rrr}
 \\[-7pt] \hline\\[-7pt]
 Complex&  & \multicolumn{2}{c}{Carina$\qquad\qquad$}  & \multicolumn{3}{c}{Cygnus}  \\
  \\[-7pt] \hline\\[-7pt]
 $t$ & $V_R\quad$ & $V_T\quad$ & $n\;$  & $V_R\quad\;$ & $V_T\quad\;$ & $n\;$   \\
 Gyr & km s$^{-1}$ & km s$^{-1}$ &  & km s$^{-1}$ & km s$^{-1}$ &   \\
\\[-7pt] \hline\\[-7pt]
0.7 & $ 6.1\pm0.1$ & $-2.5\pm0.2$ & 282 & $ 8.2\pm0.2$ & $ 9.0\pm0.1$ & 217  \\
0.9 & $ 6.3\pm0.1$ & $ 0.6\pm0.3$ & 326 & $ 7.1\pm0.4$ & $ 8.9\pm0.3$ & 153  \\
1.1 & $ 5.8\pm0.1$ & $ 3.0\pm0.2$ & 345 & $ 6.2\pm0.4$ & $ 6.7\pm0.3$ & 110  \\
1.3 & $ 5.9\pm0.2$ & $ 2.2\pm0.2$ & 320 & $ 7.4\pm0.4$ & $ 4.1\pm0.3$ & 106  \\
1.5 & $ 5.9\pm0.1$ & $ 0.2\pm0.2$ & 304 & $ 8.9\pm0.3$ & $ 3.0\pm0.2$ & 103  \\
1.7 & $ 6.5\pm0.1$ & $-1.1\pm0.1$ & 291 & $ 9.4\pm0.4$ & $ 2.9\pm0.1$ & 100  \\
1.9 & $ 6.7\pm0.2$ & $-1.2\pm0.2$ & 282 & $ 9.5\pm0.3$ & $ 2.5\pm0.2$ &  96  \\
\\[-7pt] \hline\\[-7pt]
Observations & $-6.2\pm2.6$ & $+6.2\pm2.8$ & &   $-4.3\pm1.3$ & $-10.3\pm1.4$  \\
    \\[-7pt] \hline\\[-7pt]
  \end{tabular}
 \label{mass}
\end{table*}

\subsection{Position angle of the Sun with respect to the bar major
axis}

\begin{figure*}
\resizebox{\hsize}{!}{\includegraphics{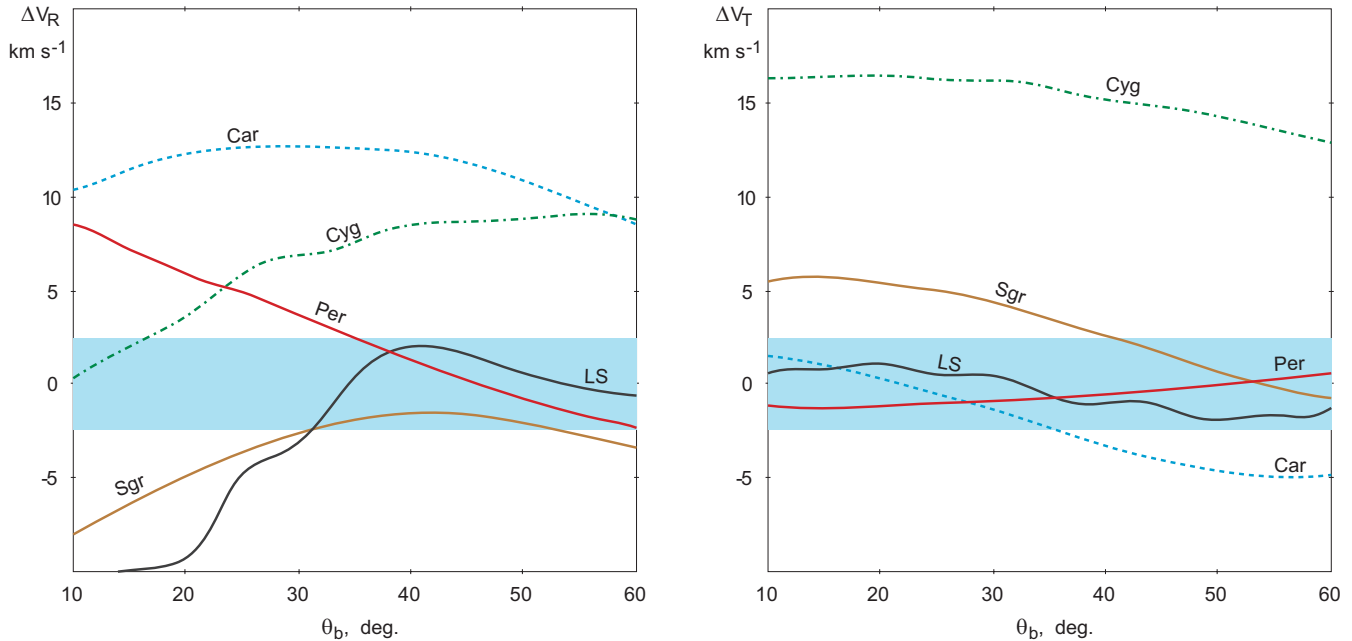}}
\caption{Dependence of the differences between model and observed
velocities, $\Delta V_R$ and $\Delta V_T$, on the position angle
$\theta_b$ of the bar calculated for five stellar-gas complexes:
Sagittarius, Carina, Cygnus, Local System (LS) and Perseus.  The
left and right panels show variations of the radial and azimuthal,
$V_R$ and $V_T$, residual velocities, respectively. The gray
strips (colored blue in electronic edition)  display the
permissable intervals of deviations between model and observed
velocities which are adopted to be $\pm2.5$ km s$^{-1}$. If a
curve indicating values of $\Delta V_R$ or $\Delta V_T$ in some
complex lies inside the strip then model and observed velocities
are consistent within the errors there.} \label{vel_theta}
\end{figure*}

In the previous analysis, the value of the  position angle of the
Sun with respect to the bar major axis  was adopted to be
$\theta_b=45^\circ$. This section  gives some rationale of that
choice. Figure~\ref{vel_theta}  shows the dependence of the
difference between  model and observed residual velocities,
$\Delta V_R$ and $\Delta V_T$:

\begin{equation}
 \Delta V_R= V_{R \; mod} - V_{R \; obs},
\label{delta VR}
\end{equation}
\begin{equation}
 \Delta V_T= V_{T \; mod} - V_{T \; obs}
\label{delta VT}
\end{equation}

\noindent on the position angle $\theta_b$. The model residual
velocities are determined as average residual velocities in model
1 at the time interval 1.4--1.6 Gyr.  The strips show the
intervals of permissable deviations between model and observed
velocities which are adopted to be $\pm2.5$ km s$^{-1}$
representing the average uncertainty in determination of observed
velocities (Table 1). If a curve indicating values of $\Delta V_R$
or $\Delta V_T$ in some complex lies inside the corresponding
strip then  the model and observed velocities are consistent
within the errors there.

Figure~\ref{vel_theta} (left panel) demonstrates that model and
observed velocities $V_R$  in the Sagittarius, LS and Perseus
stellar-gas complexes  agree within the errors under the position
angle $\theta_b$ lying in the range 33--52$^\circ$. Note that the
best agreement between model and observed velocities $V_R$  in the
Sagittarius complex corresponds to  $\theta_b \approx 45^\circ$.
The Carina and Cygnus complexes show a large discrepancy between
model and observed velocities for all values of $\theta_b$ from
the interval considered.

The most interesting feature in variations of  the azimuthal
velocity $V_T$ concerns the Sagittarius complex.
Figure~\ref{vel_theta} (right panel) indicates that model and
observed velocities $V_T$ in the Sagittarius complex are
consistent within the errors for $\theta_b > 40^\circ$. On the
contrary, the curve built for the Carina complex suggests that
model and observed velocities agree   for $\theta_b < 40^\circ$
there. Note that model velocities $V_T$ in the Local System and
Perseus complexes aren't sensitive to the choice of $\theta_b$ and
agree with observed velocities for any $\theta_b$ from the
interval considered.

Thus, the model and observed velocities, $V_R$ and $V_T$, in the
three stellar-gas complexes (Sagittarius, LS and Perseus) agree
within the errors for the position angle $\theta_b$ lying at the
interval 40--52$^\circ$.

\subsection{Surface-density profiles}

\begin{figure*}
\resizebox{\hsize}{!}{\includegraphics{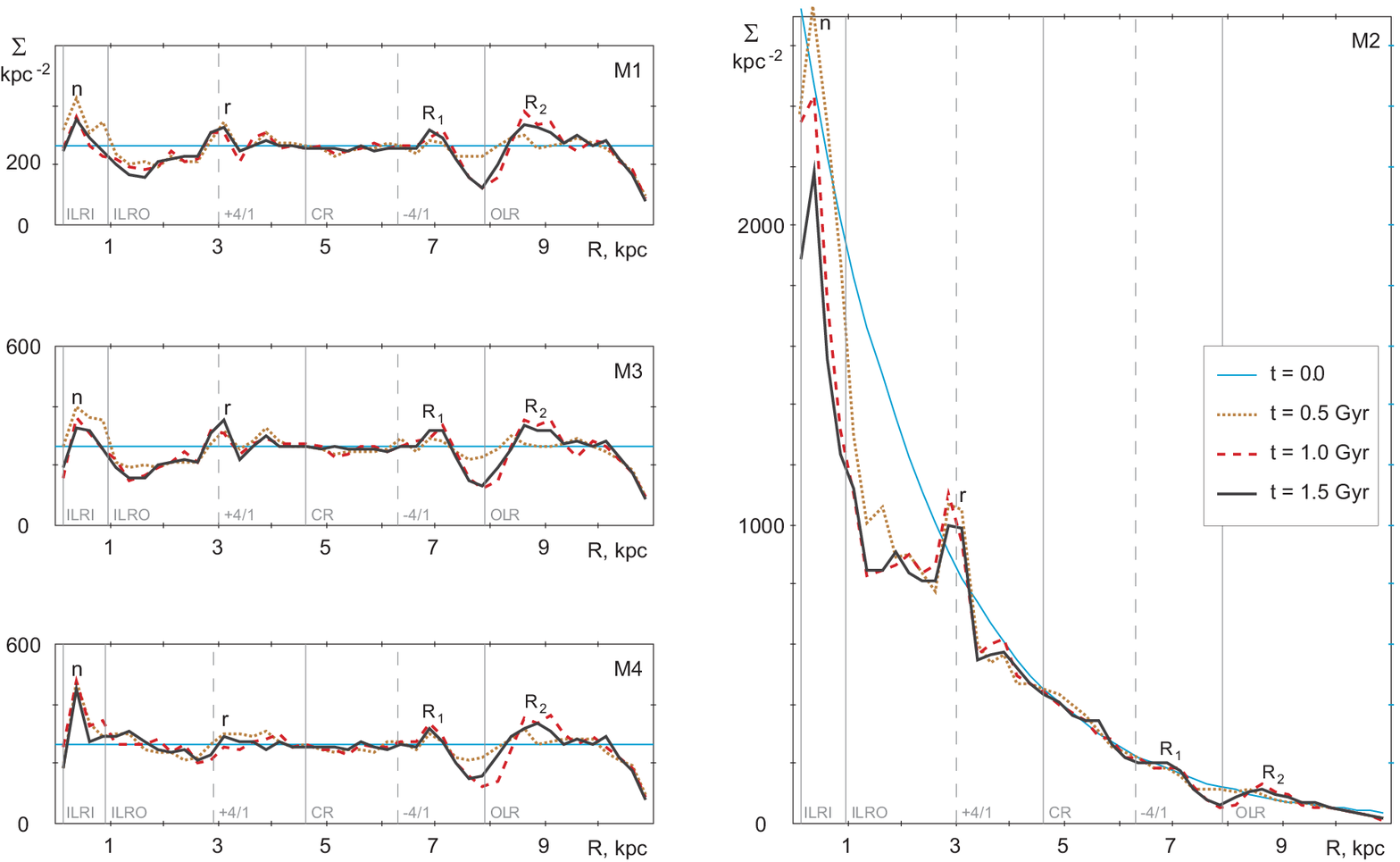}}
\caption{Profiles of the surface density $\Sigma$ built  for the
distribution of model particles (gas+OB) in models 1--4 at several
time moments: $t=0.0$, 0.5, 1.0 and 1.5 Gyr. Density maxima
related to the resonance rings are designated by letters: $n$ --
nuclear rings, $r$ -- inner rings, $R_1$ and $R_2$ -- outer rings.
The locations of the resonances are also indicated. The profile
built for model 2 exhibits a larger range of density variations
but the scale is the same in all frames. We can see that the
nuclear  rings achieve  maximum density at the time moment $t=0.5$
Gyr while the outer rings get maximum $\Sigma$ at the interval
$t=1.0\textrm{--}1.5$ Gyr. The surface density excesses above the
background are nearly the same in two outer rings but the rings
$R_2$ are nearly twice wider than $R_1$ in all models. This
suggests that the rings $R_2$ manage to catch twice more particles
than  $R_1$.} \label{den_prof}
\end{figure*}

The formation of the resonance rings can be traced by the
surface-density profiles. Figure~\ref{den_prof} shows the
variations of the surface density $\Sigma$ of model particles
(gas+OB) along the Galactocentric distance $R$  built for models
1--4 at four time moments: $t=0.0$, 0.5, 1.0 and 1.5 Gyr. These
profiles clearly indicate the positions of the resonance rings.

The nuclear rings ($n$) are forming between the two ILRs at  the
distance  of $R=0.2\textrm{--}0.9$ kpc (Fig.~\ref{den_prof}). The
surface density of the nuclear rings achieves maximum at the time
moment $t\sim 0.5$ Gyr and then starts decreasing. This process
goes most quickly in model 2 that can  be due to the largest
frequency of collisions there.

The inner rings ($r$) are growing at the distance of
$R=3.0\textrm{--}3.3$ kpc which  is a bit larger  than that of the
4/1 resonance. Note that an inner ring is practically absent in
model 4 -- we can see only a small density enhancement at $t=0.5$
Gyr there. Interestingly, the conspicuous diamond-shape structures
inside the Ferrers ellipsoid visible in many frames of
Figure~\ref{distrib}   at the distances 1--3 kpc appears to lie in
the region with the  reduced surface density
(Fig.~\ref{den_prof}).

The outer rings, $R_1$ and $R_2$, are emerging at the distances
6.7--7.3 and 8.5--9.3 kpc, respectively (Fig.~\ref{den_prof}).
They  get maximum $\Sigma$ at the interval $t=1.0\textrm{--}1.5$
Gyr, though the rings $R_2$ grow a bit slower.  The surface
density enhancements above the background are nearly the same in
two outer rings. However, the rings $R_2$ are nearly twice wider
than   $R_1$ in all models what suggests that the rings $R_2$
manage to catch twice more particles than the rings $R_1$.

On the whole, the positions and  growth rate of the resonance
rings in models considered agree with the estimates obtained in
previous simulations \citep{schwarz1981, byrd1994,
buta1996,rautiainen1999, rautiainen2000,
melnikrautiainen2009,rautiainen2010}.

\subsection{Velocity dispersion}

The velocity perturbations from the bar give rise to both
systematic motions and  velocity dispersions. To separate the
random and systematic velocities we divide model discs into annuli
of 0.5-kpc width and then make a partition of every annulus into
cells  of $\sim$0.5-kpc length in azimuthal direction. Different
annuli contain different numbers of cells. The velocities of model
particles inside every cell are assumed to obey a linear law:

\begin{equation}
V_R= V_{1} + A_1 (R-R_c) + B_1(\theta-\theta_c)+ \xi,
\label{dvr}\end{equation}
\begin{equation}
V_T= V_{2} + A_2 (R-R_c) + B_2 (\theta-\theta_c)+ \eta,
\label{dvt}
\end{equation}

\noindent where $R_c$ and $\theta_c$ are the Galactocentric radius
and  Galactocentric angle of the center of a cell, $V_{1}$ and
$V_{2}$ are the average velocities of model particles (gas+OB) in
the cell in the radial and azimuthal directions, respectively; the
parameters $A_1$, $B_1$, $A_2$ and $B_2$ describe the changes of
systematic velocities in radial and azimuthal directions, while
values $\xi$ and $\eta$ characterize the random deviations from
the linear law. In the first approximation, the standard
deviations of values $\xi$ and $\eta$ in every cell represent the
velocity dispersions in radial and azimuthal directions,
$\sigma_R$ and $\sigma_T$, respectively. The average values of
$\sigma_R$ and $\sigma_T$ calculated for all cells located in the
same annulus give us a smooth distribution of the velocity
dispersion along the Galactocentric radius.

\begin{figure}
\resizebox{8.24 cm}{!}{\includegraphics{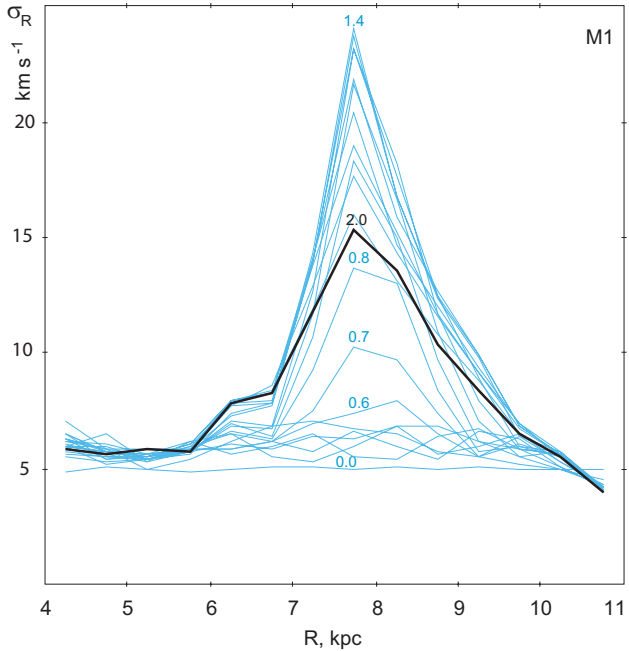}}
\caption{Dependence of  the velocity dispersion $\sigma_R$  on the
galactocentric radius $R$ in model 1 at  time moments $t$ from
$t=0$ to 2.0 Gyr every 0.1 Gyr time interval. The profile related
to $t=2.0$ Gyr is distinguished by the thick black line while
other profiles are depicted by the thin gray lines (colored blue
in electronic edition). Numbers near some profiles indicate the
time moments in Gyr. The velocity dispersion $\sigma_R$ achieves
maximum at the radius of $\sim8$ kpc at the time moment $t=1.4$
Gyr but then it starts decreasing.  All values of $\sigma_R$
located above the   value of 15 km s$^{-1}$  can be considered as
overestimated due to contribution of systematic radial velocities.
} \label{disper}
\end{figure}

Figure~\ref{disper}  shows the changes of the velocity dispersion
$\sigma_R$ along the Galactocentric distance $R$ in model 1 at
different time moments. We can see  the fast growth of $\sigma_R$
at the time interval 0.5--1.4 Gyr with  maximal value of 23 km
s$^{-1}$ being achieved at the radius of $\sim8$ kpc, but then
$\sigma_R$ declines by 15 km s$^{-1}$. All models demonstrate the
similar growth and decline in $\sigma_R$. Probably, it is the
process of the formation of the outer rings  that is responsible
for the extra increase of the velocity dispersion $\sigma_R$ at
the time interval 1.0--1.4 Gyr. We merely cannot separate properly
systematic and random motions during this process.  The drop of
$\sigma_R$ by the end of simulation is due to decreasing
systematic motions which decline especially fast in the ring $R_2$
(see variations of $V_R$ in the Perseus region,
Figure~\ref{vel_comp}, Table~5). Generally, the value of 15 km
s$^{-1}$ can be considered as the upper estimate of $\sigma_R$ at
the radius of the OLR.

Note that Figure~\ref{disper} exhibits  the velocity dispersion at
the interval of  Galactocentric distances from 4 to 11 kpc only.
In the central region the velocity dispersion achieves
considerably higher values. For example, at the distance of  the
nuclear ring, $R\approx 0.5$ kpc, $\sigma_R$  reaches $\sim100$ km
s$^{-1}$.

The velocity dispersion $\sigma_T$ is growing at the time interval
0.5--1.4 Gyr and amounts to 10 km s$^{-1}$, which is nearly twice
smaller than maximum of $\sigma_R$,  but then $\sigma_T$ decreases
by the value of 7 km s$^{-1}$.

\begin{figure*}
\resizebox{\hsize}{!}{\includegraphics{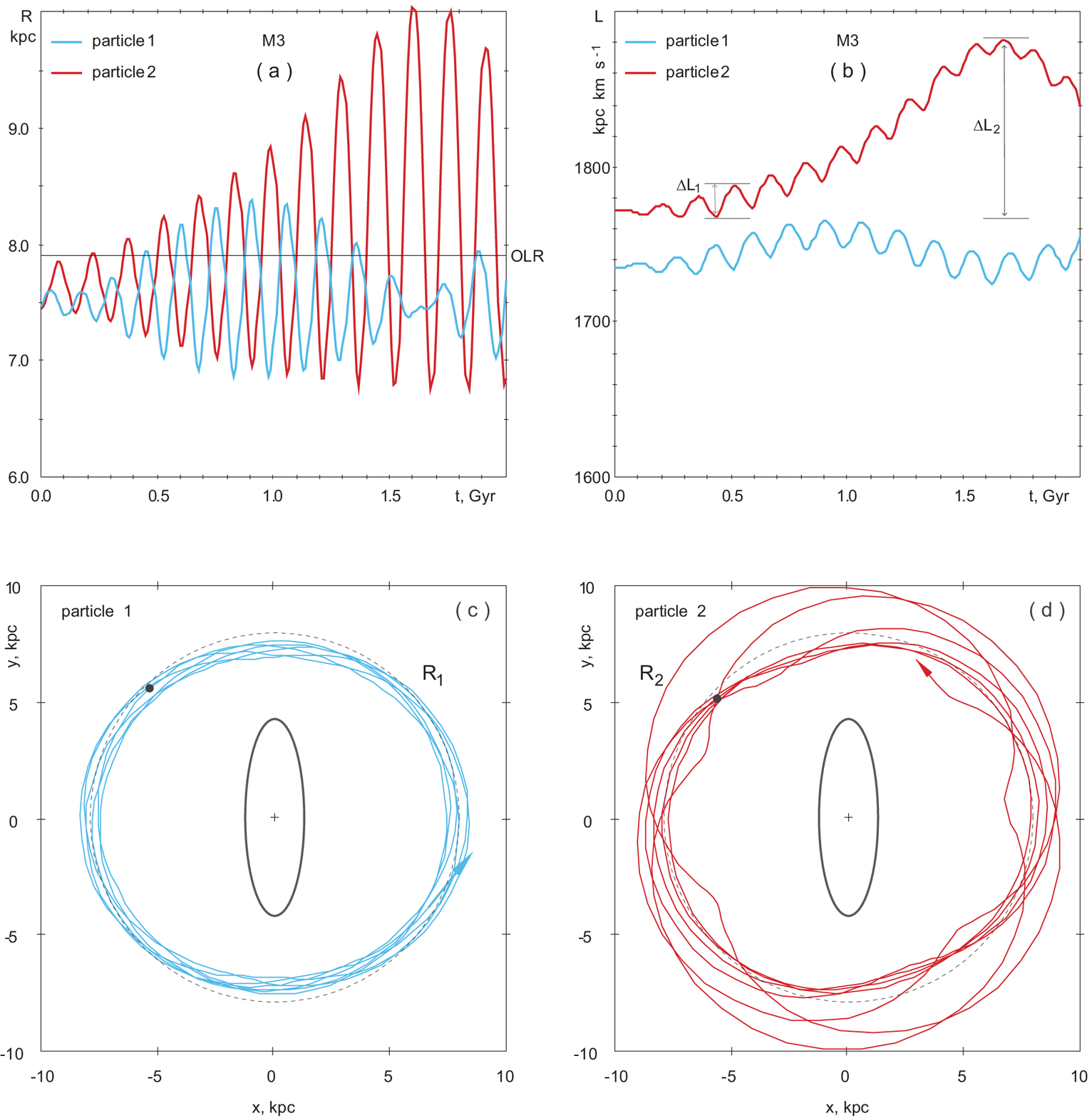}} \caption{(a)
Radial oscillations of two model particles (model 3) which appear
to lie inside the Local System at the time moment $t=1.5$ Gyr. The
oscillations  of particle 1 (colored blue in electronic edition)
and particle 2 (colored red in electronic edition) support the
outer ring $R_1$ and $R_2$, respectively.  (b) Variations of the
specific angular momentum $L$ of two chosen particles. (c) Orbit
of particle 1 in the reference frame corotating with the bar. The
ellipse indicates the position of the bar. The thin dash line
shows the radius of the OLR. A black circle in the upper left
corner points the position of the particle at $t=1.5$ Gyr. (d)
Orbit of particle 2 in the reference frame corotating with the bar
(see details  above).} \label{rad_osc}
\end{figure*}

Figure~\ref{rad_osc}(a) shows the radial oscillations of two model
particles which appear to lie inside the Local System in model 3
(one without collisions) at the time moment $t=1.5$ Gyr. Chosen
particles represent oscillations going in opposite phases. The
growth of the amplitudes evidences the resonance.
Figure~\ref{rad_osc}(b) demonstrates the variations of the
specific angular momentum $L$ and will be discussed in section
4.5. Figures~\ref{rad_osc}(c)  and \ref{rad_osc}(d) represent the
orbits of these particles in the reference frame corotating with
the bar. We can see that particle 1 supports the outer ring $R_1$
while particle 2 supports the ring $R_2$. Note that particle 1 has
the positive radial velocity $V_R$ at $t=1.5$ Gyr, when it lies
inside the Local System, while particle 2 has the negative
velocity $V_R$ at the same moment.

Figure~\ref{rad_osc}(a) indicates that  particle 1  has a maximal
amplitude of radial oscillations at $t=0.9$ Gyr approaching the
distances of 6.9 and 8.4 kpc but then the oscillations start
fading. Particle 2 deviates considerably from its initial radius,
$R=7.5$ kpc, approaching the distances of 10.0 and 6.8 kpc.
Moreover, the deviations of particle 2 in the direction away from
the Galactic center are larger than those in the opposite
direction suggesting the increase of its average distance $R$.

Figure~\ref{rad_osc}(d) shows  that the orbit of particle 2 first
isn't aligned with the bar being stretched at the angle of
$\sim45^\circ$ with respect to the bar major axis and taking an
intermediate position between the orientations of orbits in the
ring $R_1$ and $R_2$. However, the orbit has got the right
orientation being elongated  along the bar by the time $t=1.5$
Gyr.  This adjustment of orbits causes the changes in both
systematic velocities and velocity dispersions.

There is a question whether such large values of the velocity
dispersion $\sigma_R$ emerging near the OLR agree with
observations.  The velocity dispersion $\sigma_R$ achieves large
values, $\sigma_R \approx 15$ km s$^{-1}$, at the small interval
of Galactocentric distances 7.5--8.5 kpc. However, this interval
corresponds to  minimum in the distribution of the surface density
of  model particles (Fig.~\ref{den_prof}). Probably, the number of
particles with large velocity dispersion is not large. To check
that we selected OB-particles located within 3 kpc from the
adopted solar position ($R_0=7.5$ kpc, $\theta_b=45^\circ$) and
derived the parameters of the rotation curve and  velocity
dispersion from model velocities. Here we supposed that model
particles move in circular orbits in accordance with Galactic
differential rotation. The same method was applied to
observational data \citep{melnikdambis2009}. The derived rotation
curve appears to be in good agreement with the observed rotation
curve. The standard deviation $\sigma_v$ of the velocities of
OB-particles from the rotation curve computed jointly for radial
and azimuthal directions proves to be 11 km s$^{-1}$ (model 1,
$t=1.5$ Gyr).  It is a bit larger than $\sigma_v$ obtained for
observed OB-associations \citep[7--8 km s$^{-1}$,][]{melnik2017}
but still smaller than $\sigma_v$ calculated for young open
clusters \citep[15 km s$^{-1}$,][]{melnik2016} and close to
$\sigma_v$ derived for classical Cepheids \citep[10--11 km
s$^{-1}$,][]{melnik2015}.  The fraction of particles with
$|V_R|>15$ km s$^{-1}$ appears to be only 7 per cent but their
exclusion decreases the velocity dispersion to the value of
$\sigma_v=6$ km s$^{-1}$.

\subsection{Distribution of the angular momentum}

The rotation of the bar in the galactic discs causes the
redistribution of the specific angular momentum $L$:

\begin{equation}
L= \Theta R \label{L}
\end{equation}

\noindent along the Galactocentric distance $R$, where $\Theta$ is
the velocity in the azimuthal direction.

So far we have considered the kinematics near the OLR of the bar
only but in this section it makes sense to study  motions near
both Lindblad resonances: ILR and OLR. The redistribution of the
angular momentum $L$ near both Lindblad resonances seems to have
one physical reason.

\begin{figure*}
\resizebox{\hsize}{!}{\includegraphics{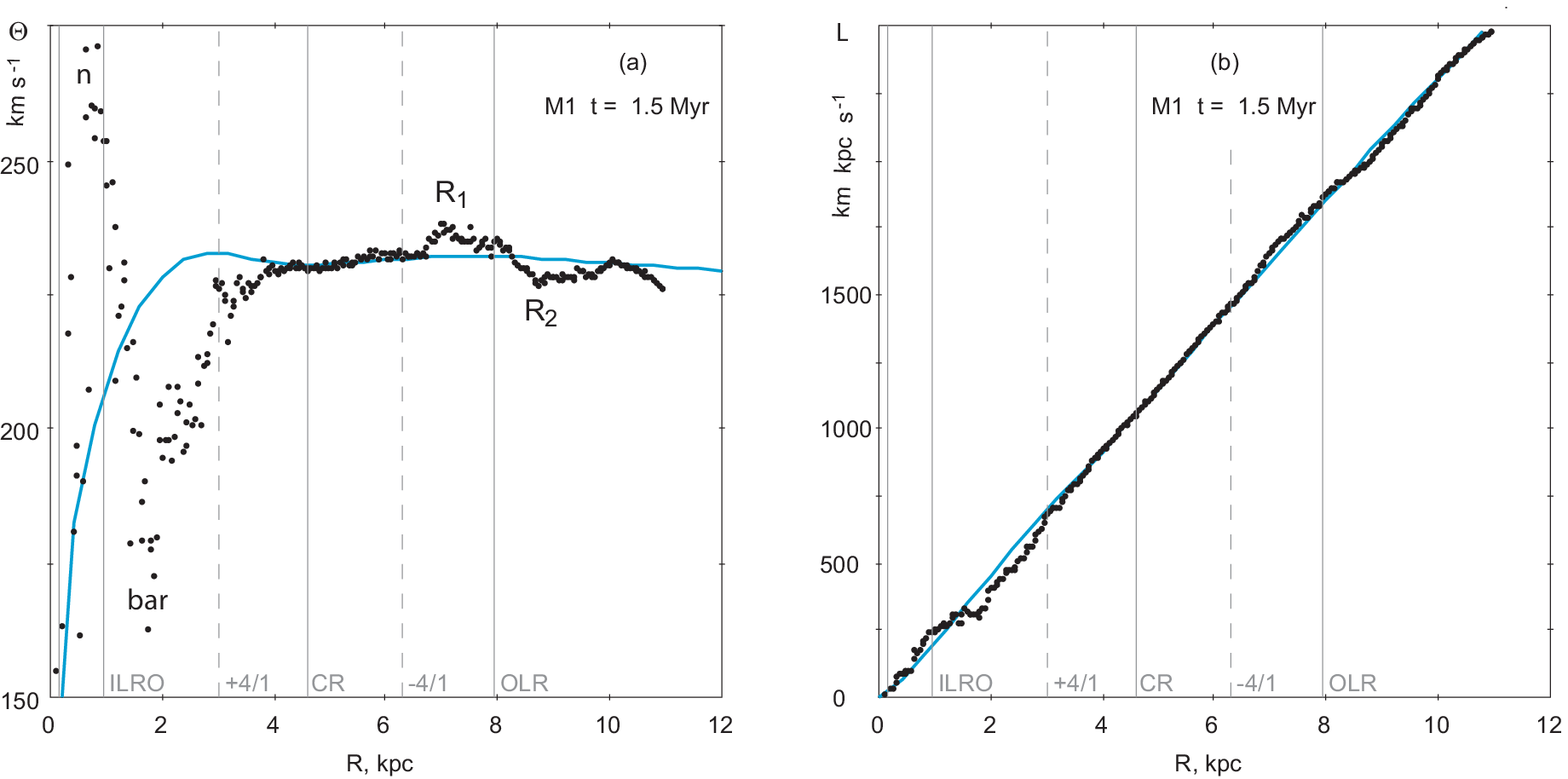}} \caption{(a)
Distribution of the azimuthal velocity $\Theta$ (black circles) of
model particles (gas+OB) averaged in thin annuli of 40-pc width
along the Galactocentic radius $R$ in model 1 at $t=1.5$ Gyr. The
gray line (colored blue in electronic edition) shows the velocity
of the rotation curve, $V_c$, which reflects the initial
distribution of $\Theta$. The vertical gray lines indicate the
positions of the resonances. We can see that the average azimuthal
velocity $\Theta$, and consequently  $L$, increases in the nuclear
region (n) and in the $R_1$-region while  $\Theta$ and $L$
decrease in the bar region and in the $R_2$-region. (b)
Distribution of the specific angular momentum $L$ (circles) of
model particles averaged in thin annuli of 40-pc width along the
distance $R$. The gray line (colored blue in electronic edition)
indicates the initial distribution of $L$. The most significant
changes of $L$ occur in the bar region. } \label{AM}
\end{figure*}

Figure~\ref{AM}(a) shows the distribution of the azimuthal
velocities $\Theta$ of model particles (gas+OB) averaged in thin
annuli of 40-pc width along the Galactocentic distance $R$ in
model 1 at $t=1.5$ Gyr. Also shown is the velocity of the rotation
curve, $V_c$, which reflects the initial distribution of $\Theta$.
We can see that particles located near the ILR and OLR of the bar
change their  velocity $\Theta$ in a similar way forming  a hump
and a pit near the radius of the resonance. The average azimuthal
velocity $\Theta$, and consequently  $L$, increases (decreases) at
the radii a bit smaller (larger) than those of the Lindblad
resonances. In the neighborhood of  the ILR, the velocity $\Theta$
grows at the distance of the nuclear ring and decreases in the
region of most populated bar orbits. In the vicinity of the OLR,
the velocity $\Theta$ increases and decreases at the distances of
the ring $R_1$ and $R_2$, respectively. Figure~\ref{AM}(b)
demonstrates the distribution of the specific angular momentum $L$
of model particles averaged in thin  annuli. The initial
distribution of $L$ is also indicated. It is seen that the most
significant changes of $L$ occur in the bar region. Note that all
models demonstrate similar behavior.

As the bar creates  accelerations in  azimuthal direction, the
angular momentum $L$ isn't conserved in barred galaxies. However,
most of particles on their quasi-periodic orbits acquire and lose
nearly the same value  of angular momentum, $\Delta L$, during
their revolution with respect to the bar.

Figure~\ref{rad_osc}(b) presents the oscillations of the specific
angular momentum of the two particles supporting the outer ring
$R_1$ and $R_2$. We can see fast oscillations of the angular
momentum, $\Delta L_1$,  with the period of $\sim 150$ Myr
corresponding to a half of their revolution period with respect to
the bar.  The range (twice amplitude) of these changes is $\Delta
L_1\approx 20$ km kpc s$^{-1}$. Besides the fast oscillations we
can see slower ones. For example, particle 2 increases its angular
momentum by the value of $\Delta L_2\approx 100$ km kpc s$^{-1}$
during the formation of the ring $R_2$ at the time interval 1--1.5
Gyr. However, both these values correspond to quite small changes
of $R$. Figure~\ref{AM}(b) demonstrates nearly linear growth of
the angular momentum $L$ with increasing $R$. Using Eq.~\ref{L}
and the value of $\Theta=232$ km s$^{-1}$ we can estimate the
variations in $R$ corresponding to $\Delta L_1$ and $\Delta L_2$
which appear to be $\Delta R_1=0.1$ and $\Delta R_2=0.4$ kpc,
respectively. Both these values are small in comparison with the
range (twice amplitude) of radial oscillations of  particle  1 and
2 equal to $\Delta R=1.5$ and 3.2 kpc, respectively
(Fig.~\ref{rad_osc}a). Thus, model particles show only small
variations of $L$  during their radial oscillations.

The resonance amplifies epicyclic motions and throws particles to
the distances corresponding to  larger changes of their angular
momenta than  $\Delta L_1$ and $\Delta L_2$ received from the bar.
Particles from smaller distances $R$ having smaller angular
momenta $L$ can come to larger distances at which particles
initially have larger $L$ and vice versa. So the average value of
the azimuthal velocity $\Theta$ and $L$ decreases (increases)  at
the radii a bit larger (smaller) than those of the Lindblad
resonances.

Probably, the redistribution of $L$ near the Lindblad resonances
of the bar is due to the existence of elongated periodic orbits
which catch a lot of particles from  nearby space. The residual
azimuthal velocities $V_T$ are directed in the opposite senses at
the apocenters (outermost points) and pericenters (innemost
points) of periodic orbits.

\begin{figure}
\resizebox{\hsize}{!}{\includegraphics{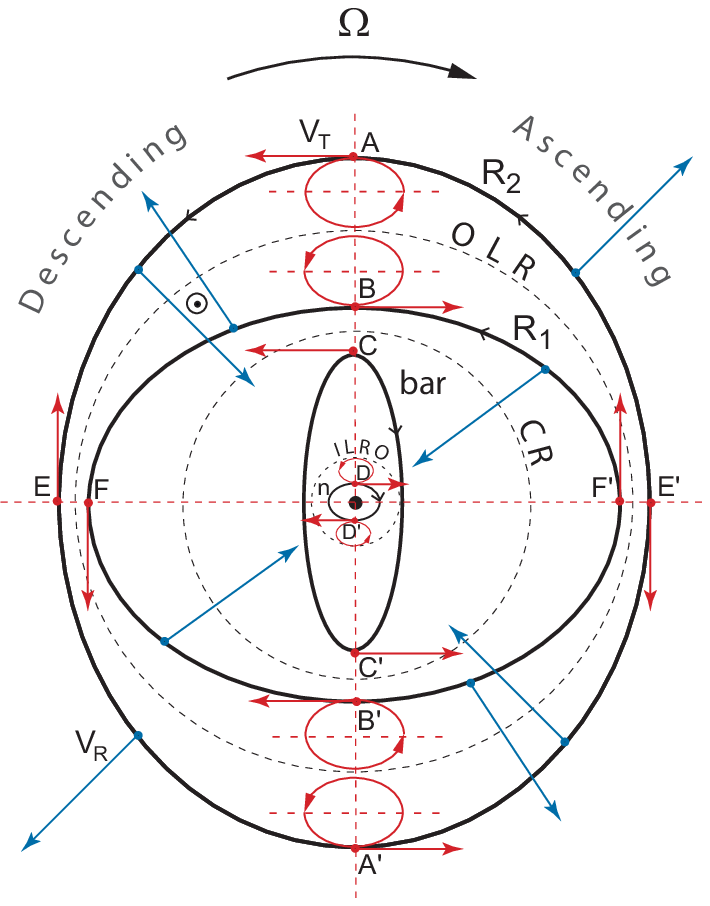}}
\caption{Schematic description of  epicyclic motions at different
points of periodic orbits supporting the nuclear ring ($n$),  bar,
outer ring $R_1$ and $R_2$.  The additional velocity $V_T$ due to
epicyclic motions (colored red in electronic edition) is directed
in the sense opposite that of galactic rotation at the apocenters
($A$, $A'$, $F$, $F'$, $C$ and $C'$) of periodic orbits and in the
sense of galactic rotation  at the pericenters ($E$, $E'$, $B$,
$B'$, $D$ and $D'$). The radial velocity $V_R$ (colored blue in
electronic edition) gets extreme values at the points lying at
about $\pm 45^\circ$-angles with respect to the bar axes. The
Galaxy rotates clockwise but in the reference frame corotating
with the bar objects located outside the CR, including those
related to the outer rings, are moving counterclockwise. The Sun
is supposed to lie at $\theta_b=45^\circ$ with respect to the bar
major axis near the descending segment ($V_R<0$) of the ring
$R_2$.} \label{schema}
\end{figure}

Figure~\ref{schema} shows the directions of the residual
velocities at different points  of periodic orbits supporting the
nuclear ring,  bar and outer rings. The additional (residual)
azimuthal velocity $V_T$ is directed in the sense opposite that of
galactic rotation ($V_T<0$) at the apocenters ($A$, $A'$, $F$,
$F'$, $C$ and $C'$) of elongated periodic orbits while $V_T$ is
directed in the sense of galactic rotation ($V_T>0$) at the
pericenters ($E$, $E'$, $B$, $B'$, $D$ and $D'$). The radial
velocity $V_R$ gets its extreme values at the points lying at
about $\pm 45^\circ$-angles with respect to the bar axes.

Probably, the tuning of epicyclic motions (Fig.~\ref{schema})
causes the appearance of annuli with deficiency and excess of
angular momentum $L$. These annuli must be located at some
distances away from the Lindblad resonances, because precisely at
the radii of the resonances, there are  both pericenters  and
apocenters of  periodic orbits oriented perpendicular to each
other there. For example, we can see that the apocenters ($F$ and
$F'$) and pericenters ($E$ and $E'$)  of periodic orbits existing
near the OLR are located practically at the radius of the OLR, so
the average value of the azimuthal velocity $\Theta$ must be close
to that of the rotation curve there. But at some distances away
from the Lindblad resonances there is nothing to compensate the
systematic changes in the azimuthal velocity. The deficiency of
$L$ ($V_T<0$) corresponds to the apocenters ($A$, $A'$, $C$ and
$C'$) of periodic orbits oriented along the bar while the excess
of $L$ ($V_T>0$) occurs at the pericenters ($B$, $B'$, $D$ and
$D'$) of periodic orbits elongated perpendicular to the bar. Thus,
the redistribution of $L$ along the radius is caused by the
existence of two types of stable periodic orbits elongated
perpendicular to each other near the Lindblad resonances of the
bar.

Let us imagine the motions of two particles located near the
points E and F at some moment and call them, for simplicity,
particle E and F, respectively (Fig.~\ref{schema}). Due to
galactic differential rotation, particle E lying  at a bit larger
$R$ must rotate with  a bit smaller angular velocity $\Omega$ than
particle F. So particle E must drift counterclockwise in the
azimuthal direction with respect to particle F.  However, the
epicyclic motions adjusted by the resonance can slow down or even
change the direction of this drift. The velocity $V_T$ at the
point E is directed in such a way to increase $\Omega$ while $V_T$
at the point F must decrease $\Omega$. Thus, the resonance can
cause the rotation of particle E and F with the same angular
velocity for some time period. This co-rotation doesn't affect
velocities of test particles in models without self-gravity, as it
is in the case considered. But if self-gravity is included, then
this co-motion can create favorite conditions for the growth of
overdensities \citep{julian1966,toomre1981,sellwood1991} and the
formation of slow modes
\citep{rautiainen2000,rautiainen2010,melnik2013}.

\section{Conclusions}

We  studied models with analytical Ferrers  bars and  compared
velocities of model particles with the observed velocities of
OB-associations. Two power indexes in the density distribution
inside the Ferrers ellipsoids were considered: $n=2$ (models 1--3)
and $n=1$ (model 4). The initial surface-density distribution of
model particles is exponential in model 2 and uniform in other
models. Model  3 doesn't include collisions but in other models
particles can collide with each other inelastically.

All models considered  can reproduce the observed residual
velocities (those after subtraction of  the velocities due to the
rotation curve and the solar motion towards the apex) of
OB-associations in the Sagittarius, Local System and Perseus
stellar-gas complexes. There are a lot of moments at the time
interval 1--2 Gyr after the start of simulations when model and
observed velocities agree  within the errors
(Fig.~\ref{vel_comp}).

The success in reproduction of the velocities in the Local System
is due to the large velocity dispersion of model particles which
weakens the resonance effects by producing smaller systematic
velocity changes.

The model  and observed residual velocities in the Sagittarius,
Local System  and Perseus stellar-gas complexes agree within the
errors under the solar position angle
$\theta_b=40\textrm{--}52^\circ$ (Fig.~\ref{vel_theta}).

The angular velocity of the bar is adopted to be $\Omega_b=50$ km
s$^{-1}$ kpc$^{-1}$ which corresponds to the location of the OLR
of the bar 0.4 kpc outside   the  solar circle, $R_{OLR}=R_0+0.4$
kpc. The uncertainty in determination of $\Omega_b$ is less than
$\pm2$ km s$^{-1}$ kpc$^{-1}$.

Model galaxies form  nuclear, inner and  outer resonance rings.
The nuclear rings  appear between the two ILRs at the distance
$\sim 0.5$ kpc from the center. The inner rings are growing at the
radius of $\sim 3.3$ kpc which is a bit larger than that of the
4/1 resonance. The outer rings, $R_1$ and $R_2$, are forming at
the radii of $\sim 7.0$ and $\sim 8.8$ kpc, respectively. The
surface density excess is  nearly the same in two outer rings.
However, the rings $R_2$ are nearly twice wider than $R_1$ in all
models what means  that the rings $R_2$ manage to catch twice more
particles than $R_1$  (Fig.~\ref{den_prof}).

The dispersion of  radial velocities, $\sigma_R$, never drops
below 5 km s$^{-1}$ in  models considered. It shows conspicuous
growth at the radius of the OLR getting maximal value of 23 km
s$^{-1}$ at 1.4 Gyr but then declines by 15 km s$^{-1}$.  The
extra growth of the velocity dispersion near the OLR seems to be
connected with the difficulty in separation between  systematic
and random motions during the formation of the outer ring $R_2$
(Fig.~\ref{disper}).

Model particles demonstrate the redistribution of the specific
angular momentum $L$  near the ILR and OLR of the bar
(Fig.~\ref{AM}). The average value of the azimuthal velocity
$\Theta$ and consequently $L$ increases (decreases)  at the radii
a bit smaller (larger) than those of the Lindblad resonances. The
most significant changes of $L$ occur in the bar region. Probably,
the redistribution of $L$ along the radius is caused by the
existence of two types of stable periodic orbits elongated
perpendicular to each other near the Lindblad resonances  of the
bar (Fig.~\ref{schema}).


\section{acknowledgements}

{\small I thank  the anonymous referee for fruitful discussion. I
thank P. Rautiainen and A.~K. Dambis   for useful remarks and
suggestions. This work has made use of data from the European
Space Agency (ESA) mission {\it Gaia}
(\verb"https://www.cosmos.esa.int/gaia"), processed by the {\it
Gaia} Data Processing and Analysis Consortium (DPAC,
\verb"https://www.cosmos.esa.int/web/gaia/dpac/consortium").
Funding for the DPAC has been provided by national institutions,
in particular the institutions participating in the {\it Gaia}
Multilateral Agreement.}


\end{document}